\documentclass[pdflatex,sn-vancouver-num,referee]{sn-jnl}
\usepackage{enumitem}
\usepackage{amsmath, amssymb}
\usepackage{mathtools}
\usepackage{graphicx}
\usepackage[font=small]{subcaption}
\usepackage{nameref}
\usepackage{hyperref}
\usepackage{textgreek}
\usepackage{tabularx}
\usepackage{multirow}
\usepackage{booktabs}
\usepackage{bm}
\usepackage{lineno}
\usepackage{xcolor}
\usepackage[section]{placeins}

\graphicspath{{./figures/}}
\newcommand{\stress}{%
    \bm{\sigma}}
\newcommand{\strain}{%
    \bm{\varepsilon}}
\newcommand{\mbf}[1]{%
				\mathbf{#1}}%
\newcommand{\ft}[1]{%
	\mathcal{F}\left[#1\right]}%
\newcommand{\ift}[1]{%
	\mathcal{F}^{-1}\left[#1\right]}%
\newcommand{\specialcell}[2][c]{%
  \begin{tabular}[#1]{@{}c@{}}#2\end{tabular}}

\begin{document}
\title{Probabilistic calibration of crystal plasticity material models with synthetic global and local data}
\author*[1]{\fnm{Joshua D.} \sur{Pribe}}\email{joshua.pribe@nasa.gov}
\author[2]{\fnm{Patrick E.} \sur{Leser}}
\author[3]{\fnm{Saikumar R.} \sur{Yeratapally}}
\author[2]{\fnm{George} \sur{Weber}}

\affil[1]{\orgname{Analytical Mechanics Associates}, \orgaddress{\city{Hampton}, \state{VA}, \country{USA}}}
\affil[2]{\orgname{NASA Langley Research Center}, \orgaddress{\city{Hampton}, \state{VA}, \country{USA}}}
\affil[3]{\orgname{Science and Technology Corporation}, \orgaddress{\city{Hampton}, \state{VA}, \country{USA}}}

\abstract{
Calibrating full-field crystal plasticity (CP) models using global stress-strain curves alone often produces non-unique parametrizations: multiple parameter sets can predict the same global behavior but different local, grain-scale behavior.
Incorporating local calibration data can mitigate uniqueness issues, but this typically requires specialized experiments like high-energy X-ray diffraction microscopy.
The computational expense of full-field simulations also prevents uncertainty quantification with sampling-based probabilistic calibration algorithms like Markov chain Monte Carlo.
In this study, a two-stage probabilistic calibration procedure is developed that uses both global and local data and balances the efficiency of a surrogate model with the accuracy of a full-field CP model.
To enable tractable Bayesian inference via sampling from the full-field model, an efficient, parallelized sequential Monte Carlo (SMC) algorithm is used to perform the calibration, which involves more than \mathversion{normal}$100,000$ CP simulations.
Synthetic measurement data is used in the calibration to assess uncertainty and accuracy of the parameter posterior distributions relative to a ``ground-truth'' simulation with a microstructure representative of Inconel 718.
The two-stage calibration is shown to be robust to limited and noisy local data. 
Posterior distributions and computational costs are compared between the two-stage calibration and a one-stage calibration using only the full-field model, which is also enabled by SMC, as well as an efficient but biased one-stage calibration using a surrogate model.
Overall, including local measurement data is shown to reduce parameter uncertainty, while joint distributions of the calibrated parameters highlight important considerations in choosing constitutive models and calibration data, including challenges resulting from correlated parameters.
}


\maketitle

\section*{Introduction}
\label{sec:intro}
Full-field crystal plasticity (CP) models have become powerful tools for predicting deformation and failure in polycrystalline alloys, including local phenomena like fatigue crack initiation in \cite{mcdowell_microstructure-sensitive_2010,roters_overview_2010,lebensohn_spectral_2020}.
Although CP models provide a physical, mechanistic basis for plasticity, error and uncertainty in the model form and parameters can limit predictive accuracy at local length scales \cite{pokharel_polycrystal_2014,lim_quantitative_2025,proudhon_situ_2020}.
For complex phenomenological models with many parameters, independently calibrating all model parameters is often extremely challenging due to limited measurement data \cite{ghosh_non-deterministic_2020,ashraf_history_2023}.

Traditional calibration approaches for CP models rely on macroscale or global measurements, such as uniaxial stress-strain curves \cite{rovinelli_assessing_2017,bandyopadhyay_uncertainty_2019,herrera-solaz_inverse_2014}.
However, many studies have demonstrated that multiple CP parameter sets may be consistent with the global stress-strain curve, yet predict widely varying responses for single crystals \cite{sangid_validation_2014,dindarlou_substructure-sensitive_2022,shimanek_simultaneous_2025}.
In other words, calibrating a model using the global stress-strain curve alone results in a lack of unique and identifiable parameters.
Non-unique parametrizations in turn can cause the model to be unreliable for predicting local quantities of interest like intragranular stress concentrations \cite{ashraf_history_2023,rezwan_coupled_2025}.
Therefore, obtaining calibration and validation data at multiple scales, including the scale of individual grains, is necessary to successfully predict extrema of local fields \cite{pokharel_polycrystal_2014,sangid_coupling_2020,joshi_integrated_2025,proudhon_situ_2020}.

High-fidelity measurement techniques have begun to provide access to local measurement data.
High-energy X-ray diffraction microscopy (HEDM) can be used to estimate grain-average elastic strains and infer stresses in three dimensions \cite{shade_exploring_2019,obstalecki_quantitative_2014}.
HEDM has been used to enhance mechanistic understanding \cite{menasche_four-dimensional_2023,pagan_modeling_2017} and validate CP predictions at the scale of individual grains within a polycrystal \cite{menasche_afrl_2021,mackey_grain_2023,prithivirajan_direct_2021} and across different loading regimes \cite{yeratapally_discrepancy_2021,tari_validation_2018}.
High-resolution electron backscatter diffraction (HR-EBSD) provides similar data, but only on the surface of a specimen.
During an in-situ loading experiment, HR-EBSD combined with digital image correlation (DIC) can estimate local strains, dislocation densities, and crystallographic texture evolution to support model calibration \cite{savage_identification_2021,choi_multi-objective_2026} and validation \cite{gilliland_concurrent_2026,zhang_crack_2016,sahu_micro-mechanisms_2024,stinville_insights_2023}.
However, local measurements from both HR-EBSD and HEDM are limited due to the time and monetary costs of operating scanning electron microscopes and especially high-energy synchrotrons \cite{shade_exploring_2019,proudhon_situ_2020,stinville_insights_2023}.
As a result, it is important to understand how many and what types of measurements to make with these techniques.

In addition to limited measurement data, high computational cost and complex parameter spaces also cause significant challenges for CP calibration procedures.
Hand-tuning the model parameters is the simplest approach \cite{cheng_macro-micro_2024,liu_strategy_2020,park_crystal_2019} but becomes more difficult as the number of parameters increases and provides minimal information about accuracy and uncertainty \cite{cardoso_coelho_efficient_2024}.
Deterministic optimization algorithms are more rigorous than hand-tuning but typically provide only point estimates of the calibration parameters and may converge to one of many local minima due to non-uniqueness in the high-dimensional parameter space, complex parameter interactions, and limited or noisy data \cite{ghosh_non-deterministic_2020,depriester_crystal_2023,cardoso_coelho_efficient_2024}.
Probabilistic methods like Bayesian inference instead provide both point estimates and uncertainty quantification (UQ) for the parameters, which can be crucial for predicting reliability \cite{fernandez-zelaia_creep_2022,whelan_uncertainty_2019}.
However, probabilistic calibrations typically use sampling methods like Markov chain Monte Carlo (MCMC) that can require many thousands of model evaluations, greatly increasing cost over deterministic methods \cite{ghosh_non-deterministic_2020}.

Existing strategies to improve the computational tractability of calibrating full-field CP models all involve significant trade-offs.
Efficient surrogate models like Gaussian processes have been used in deterministic and probabilistic calibration schemes \cite{nguyen_bayesian_2021,fernandez-zelaia_creep_2022,veasna_machine_2023} and to compare different model forms given local data \cite{venkatraman_bayesian_2022,venkatraman_bayesian_2025}.
However, the resulting calibrations can be biased due to discrepancies between the surrogate and the CP model \cite{nguyen_bayesian_2021,kucerova_efficient_2023,choi_multi-objective_2026}, and the prediction error and training costs scale with the dimensionality of the inputs (i.e., calibration parameters) and outputs (i.e., adding local measurement data) \cite{brynjarsdottir_learning_2014}.
Calibrations that rely on repeatedly evaluating the full-field CP model typically use gradient-free methods like genetic algorithms \cite{bandyopadhyay_uncertainty_2019,weber_machine_2022,sedighiani_efficient_2020,choi_multi-objective_2026} and Bayesian optimization \cite{kuhn_identifying_2022,tran_asynchronous_2023,shimanek_simultaneous_2025}.
Both of these methods reduce model evaluations relative to gradient-based optimization \cite{cardoso_coelho_efficient_2024}, but they typically return only point estimates of the calibration parameters.
Repeated genetic algorithm evaluations can provide basic UQ subject to restrictive assumptions on the parameter realizations \cite{bandyopadhyay_uncertainty_2019,choi_multi-objective_2026}.
More rigorous UQ using MCMC with full-field CP models is generally seen as computationally intractable \cite{ghosh_non-deterministic_2020,bandyopadhyay_uncertainty_2019,kucerova_efficient_2023}.

Finally, two-stage calibration approaches were used in \cite{depriester_crystal_2023} and \cite{nguyen_bayesian_2021} with the goal of obtaining informative initial guesses or prior distributions that reduce the number of required evaluations of a full-field CP model.
Both used an inexpensive model with global uniaxial data in the first stage and introduced local data in the second stage.
Depriester et al.\ \cite{depriester_crystal_2023} used a full-field CP model and deterministic optimization in the second stage but still encountered uniqueness issues and extreme computational costs (about 55 days of walltime) despite incorporating information from the inexpensive first stage.
Nguyen et al.\ \cite{nguyen_bayesian_2021} trained a Gaussian process surrogate for their CP model in the second stage on a restricted parameter space informed by the first stage output.
The surrogate model allowed for UQ through Bayesian inference but is subject to the limitations of surrogate models noted previously, and the authors also commented that hand-tuning some of the parameters was ``more straightforward'' \cite{nguyen_bayesian_2021}.

\subsubsection*{Contribution of This Study}
This study addresses two gaps in prior work.
First, the computational expense of sampling from full-field CP models has led to a reliance on restrictive assumptions and surrogate models for probabilistic calibration, which can lead to predictive bias.
Prior work using Bayesian inference to calibrate expensive mechanical models like crystal plasticity relies on low-fidelity or surrogate models to make MCMC algorithms tractable (e.g., \cite{nguyen_bayesian_2021,ghosh_non-deterministic_2020,fernandez-zelaia_creep_2022,venkatraman_bayesian_2025,kucerova_efficient_2023}).
Here, an efficient sequential Monte Carlo (SMC) algorithm is used to perform Bayesian inference on the CP parameters using many parallel model evaluations with both global and local measurement data.
The SMC parallelization enables computationally tractable UQ with the full-field CP model, addressing previous computational bottlenecks and mitigating surrogate-induced bias.
A two-stage calibration procedure is also developed to balance the efficiency of a surrogate model in the first stage with the accuracy of the full-field CP model in the second stage.

Second, while prior work has shown that local data is needed to resolve parameter uniqueness issues, it is necessary to assess how the availability and quality of this data affects the uncertainty in the calibrated parameters.
In this study, results from several calibration runs with different model assumptions and data availability are compared, including calibrations with global data only, local data with increased measurement noise, and with and without a surrogate model.
Synthetic measurement data is used in all calibrations and generated by a CP simulation with ground-truth parameters based on the nickel-base superalloy Inconel 718 \cite{stinville_multi-modal_2022,bandyopadhyay_uncertainty_2019}.
Three research questions are addressed by this part of the study:
\begin{enumerate}
	\item How does the availability of local data improve the calibration results in terms of both accuracy and uncertainty?
	\item How robust is the two-stage approach to limited or noisy local data?
	\item How do the calibration results inform how constitutive models and calibration experiments should be selected?
\end{enumerate}

\begin{figure}
	\centering
	\includegraphics[width=0.98\textwidth]{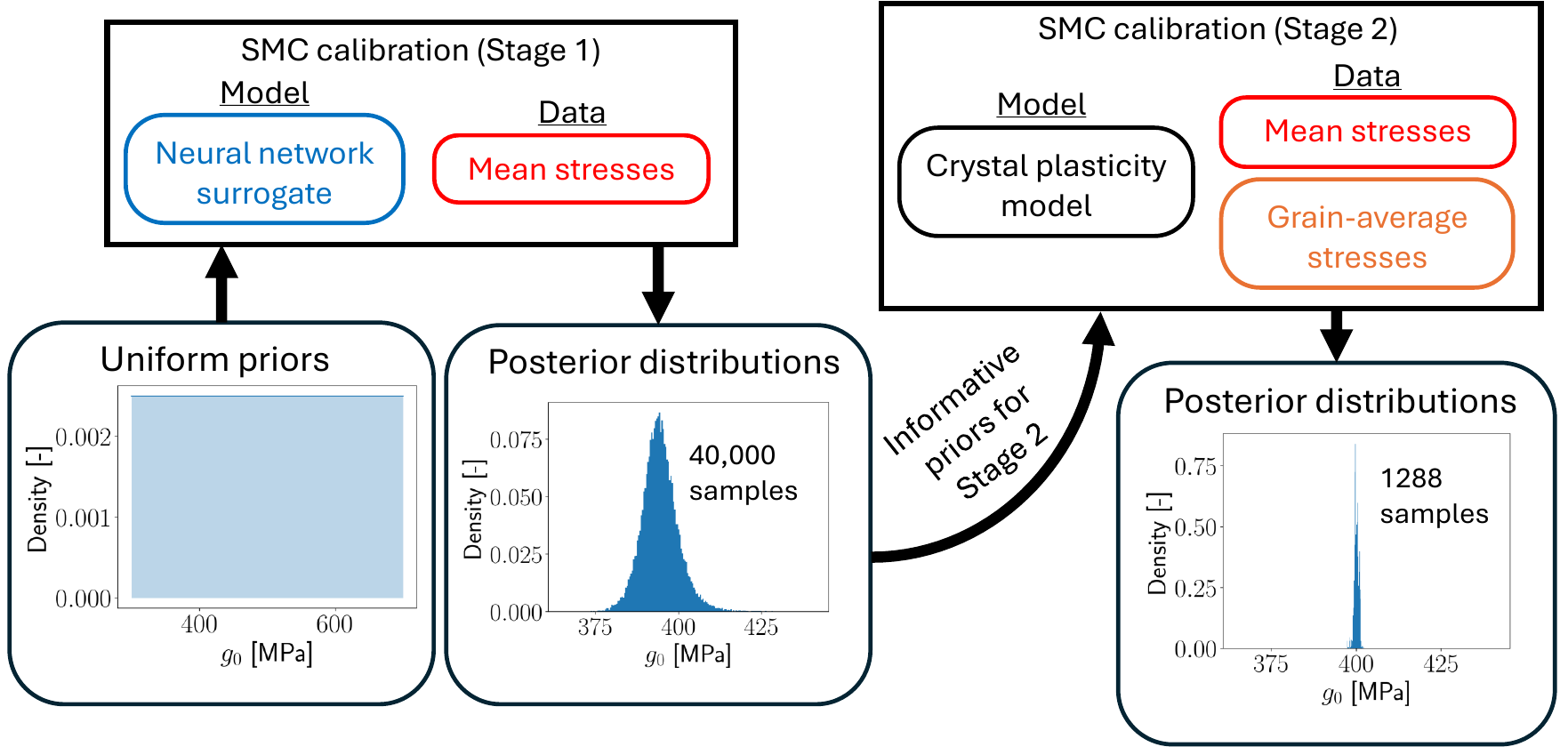}
	\caption{Workflow for the two-stage calibration procedure.
	Stage 1 uses uniform prior distributions on each parameter, the neural network surrogate model, and the synthetic mean stress data.
	Stage 2 uses priors based on a kernel density estimate of the material parameter posterior distributions from stage 1, the full CP model, and both the synthetic mean and grain-average stress data.}
	\label{fig:figure1}
\end{figure}

The overall workflow for the two-stage calibration procedure is shown in \autoref{fig:figure1}.
Stage one uses a neural network surrogate model to infer preliminary posterior distributions for the model parameters based on only the global data.
The posteriors serve as informative prior distributions for stage two, which uses the full-field CP model with both global and local data to infer the final posterior distributions for the model parameters.
The informative priors are intended to reduce the number of CP model evaluations.
%

The remainder of this study is organized as follows.
The \nameref{sec:cp} section describes the CP formulation, constitutive model, and microstructure.
The \nameref{sec:calibration} section describes the SMC algorithm, calibration parameters, synthetic measurement data, and proposed two-stage calibration approach.
The \nameref{sec:results} section first presents posterior distributions from each stage of the two-stage calibration.
The influence of limited and noisy data on the posterior distributions is then studied, and the two-stage calibration is compared with one-stage calibrations using only the surrogate and only the CP model.
The \nameref{sec:discussion} section describes advantages and limitations of the calibration approach based on the numerical results and identifies challenges and potential solutions for applying the approach to real measurement data.
Finally, the \nameref{sec:conclusions} section summarizes important contributions of this study.

\section*{Crystal Plasticity Simulations}
\label{sec:cp}
The CP simulations in this paper use the fast Fourier transform (FFT)-based small-strain Fourier-Galerkin scheme of \cite{zeman_finite_2017,de_geus_finite_2017}.
The stress and strain are discretized on a regular grid of voxels with trigonometric polynomial interpolation.
A projection tensor is defined in frequency space that enforces equilibrium and compatibility, and the resulting system of equations is solved using the Newton-Raphson method in each time increment.
Specific components of the projection tensor are used to enforce stress and strain boundary conditions\footnote{Strictly speaking, in FFT-based solvers, conventional boundary conditions cannot be enforced without modifying the algorithm, due to the inherent periodicity. Rather, the \emph{global mean} stress and strain components must be specified.} \cite{lucarini_algorithm_2019}.
Further details can be found in Appendices \ref{app:solver} and \ref{app:projection}.
The model was implemented in version 0.1 of the open-source code Materialite\footnote{Available under the Apache 2.0 license at \url{https://github.com/nasa/materialite}.} developed at NASA Langley Research Center.

\subsubsection*{Constitutive Model}
Assuming small strains with the standard additive decomposition, the stress at a point is given by
\begin{equation}
    \label{eq:additive_decomposition}
    \stress = \mathbb{D} \left( \strain - \strain_p^{(t)} - \dot{\strain}_p \Delta t \right),
\end{equation}
where $\stress$, $\strain$, and $\dot{\strain}_p$ are the stress, total strain, and plastic strain rate, respectively, at time $t + \Delta t$; $\strain_p^{(t)}$ is the plastic strain at time $t$; $\Delta t$ is the time increment, and $\mathbb{D}$ is the elasticity tensor.
A viscoplastic flow rule relates the stress to the plastic strain rate:
\begin{equation}
    \label{eq:flow}
    \dot{\strain}_p = \dot{\gamma}_0 \sum_{s=1}^{N_s} \mbf{m}_s \left( \frac{\left| \mbf{m}_s : \stress \right|}{g_s} \right)^m \operatorname{sgn} \left(\mbf{m}_s : \stress \right),
\end{equation}
where $s$ is a slip system identifier, $N_s$ is the total number of slip systems, $\dot{\gamma}_0$ is a reference strain rate, $m$ is the inverse rate sensitivity, and $\mbf{m}_s$ and $g_s$ are the Schmid tensor and critical resolved shear stress (CRSS), respectively, for slip system $s$.
The model uses the twelve $\{111\}\left<1\bar{1}0\right>$ octahedral slip systems typical of face-centered cubic materials.

For the calibration study, it was desired to use a relatively simple hardening law that still demonstrated interesting behavior and had approximate ground-truth values and intervals for each parameter in prior work \cite{bandyopadhyay_uncertainty_2019}.
To this end, a basic hardening/dynamic recovery model (similar to Armstrong-Frederick-type hardening laws) was chosen:
\begin{equation}
    \label{eq:hardening}
    \dot{g}_s = \left( H - H_d g_s\right) \dot{\Gamma}, \quad \dot{\Gamma} = \sum_{s=1}^{N_s} \left| \dot{\gamma}_s \right|, \quad g_s = g_0 + \int_0^t \dot{g}_s \mathrm{d}t,
\end{equation}
where $g_0$ is the initial CRSS on all slip systems, $H$ is the hardening coefficient, $H_d$ is the dynamic recovery coefficient, and $\Gamma$ is the accumulated slip.
The self and latent hardening coefficients are assumed to be unity.

\subsubsection*{Model Geometry and Boundary Conditions}
%
\begin{figure}
	\centering
	\includegraphics[width=0.7\textwidth]{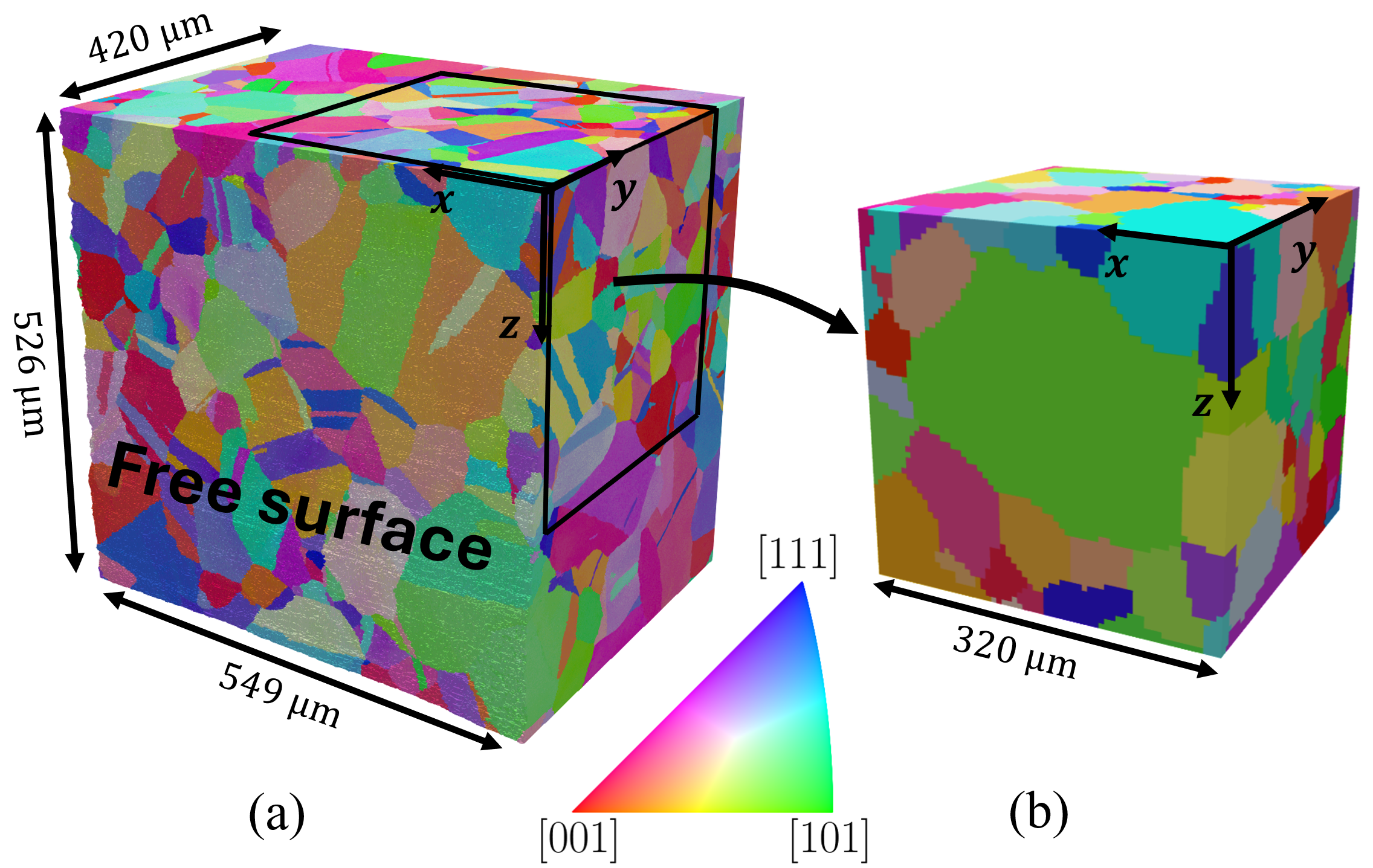}
	\caption{IPF maps of (a) the full microstructure, reconstructed from the data in \cite{stinville_multi-modal_2022}, and (b) the simplified microstructure used in the simulations.
    The reference direction is the $x$ direction for both IPF maps. The black-outlined region in (a) corresponds to the microstructure in (b).}
	\label{fig:microstructures}
\end{figure}
The microstructure for the simulations was taken from an open-source dataset \cite{stinville_multi-modal_2022}, which contains 3-D microstructural information for a 549 \textmugreek m $\times$ 526 \textmugreek m $\times$ 420 \textmugreek m sample of Inconel 718.
An inverse pole figure (IPF) map of the microstructure is shown in \autoref{fig:microstructures}a.
The microstructure has a 1-\textmugreek m voxel size, one free surface, and a large number of annealing twins.
To make the computations in this study more tractable, 30 voxels were removed from the $-y$ face to eliminate the free surface, the voxel size was coarsened to 5 \textmugreek m, twins were merged into their parent grains, and a $64 \times 64 \times 64$-voxel region, shown in \autoref{fig:microstructures}b, was extracted for use in the simulations.
Finally, a buffer region with a thickness of 6 voxels was added on the $y$ and $z$ faces.
The buffer was assigned an isotropic elastic and perfectly plastic material model, such that the microstructure approximates a smaller region within the bulk of a tensile specimen.

Fixed values for the elastic constants ($C_{11}$, $C_{12}$, and $C_{44}$) \cite{ghorbanpour_crystal_2017}, reference strain rate \cite{bandyopadhyay_uncertainty_2019}, and buffer material properties (Young's modulus, $E$; shear modulus, $\mu$; and yield stress, $\sigma_Y$) are shown in \autoref{tab:params}.
The buffer was chosen to be more compliant and have a lower yield stress than the microstructure to minimize the influence of the buffer while still providing constraint to the model.
Specifically, Young's modulus and the shear modulus were chosen as the directional stiffness and directional shear modulus of the microstructural elasticity tensor in the crystal coordinate directions.
The yield stress was chosen such that it would be below the global yield stress across all parameter combinations.
%
\begin{table}
	\centering
	\caption{Fixed material parameters.}
	\begin{tabular}{ccc}
		\toprule
		Material region & Symbol & Value \\
		\midrule
		\multirow[t]{4}{*}{Microstructure} & $C_{11}$ & 242.2 GPa \\
			\cmidrule{2-3} & $C_{12}$ & 138.9 GPa \\
			\cmidrule{2-3} & $C_{44}$ & 104.2 GPa \\
			\cmidrule{2-3} & $\dot{\gamma}_0$ & 0.004 s$^{-1}$ \\
		\midrule
		\multirow[t]{3}{*}{Buffer region} & $E$ & 140.9 GPa \\
			\cmidrule{2-3} & $\mu$ & 104.2 GPa \\
			\cmidrule{2-3} & $\sigma_Y$ & 300 MPa \\
		\bottomrule
	\end{tabular}
	\label{tab:params}
\end{table}

In each simulation, a constant mean strain rate was applied in the $x$ direction up to 1\% strain.
The mean strain rate was set equal to the reference strain rate in \autoref{tab:params}.
The mean stress for all other components was fixed to zero.
See Appendix \ref{app:projection} for more details.

\section*{Calibration Approach}
\label{sec:calibration}
\subsubsection*{Calibration as a Bayesian Inference Problem}
\label{sec:calibration_overview}
The goal of the calibration is to quantify uncertainty in a set of calibration parameters, $\bm{\theta} \in \mathbb{R}^p$, given noisy measurement data, $\mbf{y} \in \mathbb{R}^q$.
Here, $p$ is the number of parameters to be calibrated, and $q$ is the number of experimentally measured output quantities (e.g., number of points on a stress-strain curve).
A Bayesian perspective is adopted in this study, meaning the output of the calibration is a joint probability distribution over the model parameters reflecting the analyst{'}s belief about their true values after observing data.
This distribution is referred to as the posterior and, from Bayes{'} theorem, has probability density function (PDF)
\begin{equation}
    \label{eq:bayes}
    p(\bm{\theta} | \mbf{y}) = \frac{p(\mbf{y} | \bm{\theta}) p(\bm{\theta})}{Z},
\end{equation}
where $p(\mbf{y} | \bm{\theta})$ is the likelihood function, $p(\bm{\theta})$ is the prior PDF for the parameters, and $Z$ is a normalizing constant referred to as the marginal likelihood or model evidence.

The likelihood function quantifies the plausibility of observing $\mbf{y}$ if $\bm{\theta}$ were the true parameter values.
Its definition depends on the assumed relationship between the data, the model, and the noise,
\begin{equation}
    \label{eq:likelihood}
    \mbf{y} = \mathcal{M}(\bm{\theta}) + \bm{\epsilon},
\end{equation}
where $\mathcal{M}(\cdot)$ is the model and $\bm{\epsilon} \sim \mathcal{N}(0, \Sigma)$ is zero-mean Gaussian measurement noise.
The Gaussian noise assumption produces a closed-form likelihood, which enables direct evaluation of the posterior PDF up to some proportionality $p(\bm{\theta} | \mbf{y}) \propto p(\mbf{y} | \bm{\theta}) p(\bm{\theta})$ for any vector $\bm{\theta}$.
As a result, random walk algorithms like MCMC can be employed to produce samples from the posterior without knowing $Z$, as will be discussed in Section \ref{sec:smc}. 
For most cases, a constant variance was assumed for each measurement such that $\Sigma = \bm{I}\sigma_\text{noise}^2$.

\subsubsection*{Sequential Monte Carlo}
\label{sec:smc}
In a standard MCMC algorithm for sampling from the posterior distribution, each proposed sample requires an evaluation of the model to compute $p(\mbf{y} | \bm{\theta})$ prior to accepting or rejecting the sample.
Each evaluation is completed sequentially due to the Markov nature of the random walk, which is computationally intractable for the full-field CP model.
To overcome the computational burden, SMC, a parallelizable alternative to MCMC, was adopted.

The computational advantage of SMC is derived from its use of a population of independent, weighted parameter vectors called ``particles''.
The particles are evolved through a series of intermediate target distributions via successive sequential importance sampling and resampling steps.
The target distributions are defined using likelihood annealing, such that the $i^\text{th}$ target is $p_i(\bm{\theta} | \mbf{y}) \propto p(\mbf{y} | \bm{\theta})^{\phi_i} p(\bm{\theta})$.
The exponent $\phi_i \in [0,1]$ monotonically increases, such that the first target is simply the prior and the last target is the posterior.
Transitioning the particles from one target distribution to the next is referred to as an SMC step.

For each target distribution, importance weights, $w_j$, are assigned to the $j^\text{th}$ particle,
\begin{equation}
	w_j \propto \frac{p(\bm{\theta}_j)p(\mbf{y} | \bm{\theta}_j) ^{\phi_i}}{p(\bm{\theta}_j)p(\mbf{y} | \bm{\theta}_j) ^{\phi_{i-1}}} = p(\mbf{y} | \bm{\theta}_j) ^{\phi_i - \phi_{i-1}}.
\end{equation}
Resampling with replacement then occurs to move low-weight particles to high-weight regions of the parameter space.
Over the course of several SMC steps, particles will therefore be moved toward regions of high posterior probability density.
However, this resampling approach often produces a degenerate population with most particles resampled to a small subset of the original population, resulting in very few unique values of $\bm{\theta}$.
The reason is that the support of the prior distribution is typically much broader than the posterior distribution.
To mitigate degenerate populations, each SMC step includes a short MCMC run for each particle, enabling local exploration of the $i^\text{th}$ target distribution via random walk.
In summary, each SMC step involves computing the importance weights, resampling the particles, and completing an MCMC run for each particle.

SMC steps are carried out for each target until termination of the algorithm at $\phi_i=1$.
The end result is similar to MCMC in that the final particle population represents samples from the posterior distribution.
However, all particles are moved independently, meaning that evaluations of the likelihood within each short MCMC run can be parallelized across particles.
This results in significant speedup that scales well with additional computational resources.

The SMC computations in this study were completed using the open-source SMCPy code\footnote{Available at \url{https://github.com/nasa/SMCPy}.} developed at NASA Langley Research Center.
For all computations, five MCMC proposals (corresponding to five likelihood evaluations) were used per particle per SMC step.
The number of particles varied depending on computational cost and will be provided along with the relevant results.
SMCPy{'}s \texttt{AdaptiveSampler} class was used, meaning each $\phi_i$ was selected adaptively such that the resampled population met a target effective sample size, defined as the percentage of particles with meaningful importance weights.
This avoided additional hyperparameters to define the target sequence while ensuring target distributions evolved gradually enough to avoid sampling errors.
The tradeoff was that the total number of SMC steps and thus the total walltime of each calibration was not known in advance.
More details on SMC algorithms can be found in \cite{del_moral_sequential_2006,del_moral_adaptive_2012,leser_sequential_2018}.

\subsubsection*{Calibration Parameters}
\label{sec:calibration_parameters}
\begin{figure}
	\centering
	\includegraphics[width=0.45\textwidth]{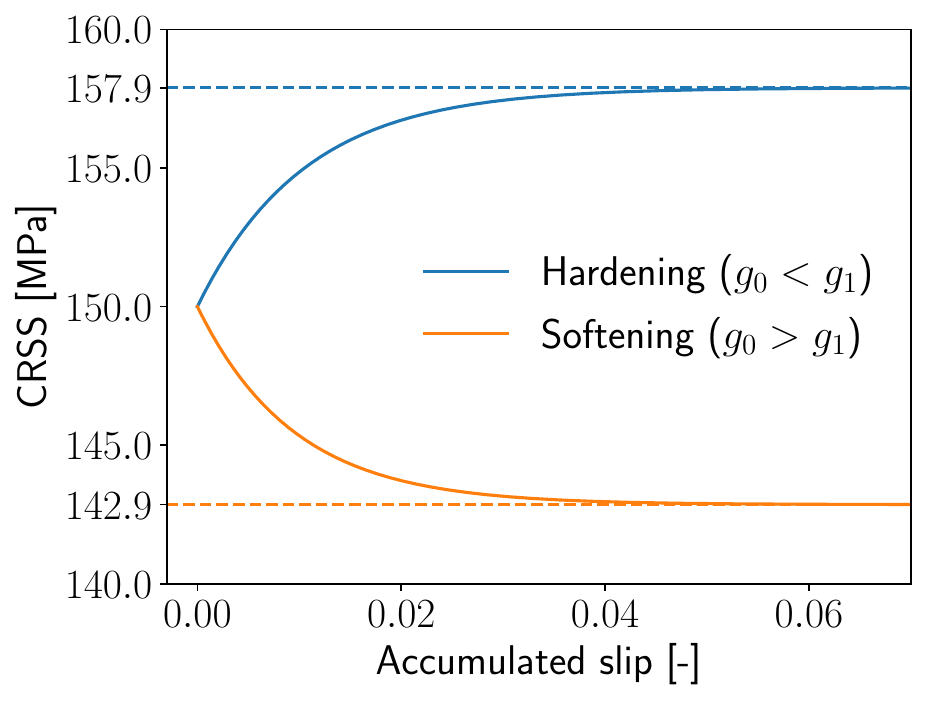}
	\caption{Examples of cases where strain hardening and strain softening occur.
    Both curves have $g_0 = 150$ MPa and $H = 15,000$ MPa.
    The blue hardening curve has $H_d = 95$, resulting in $g_1 \approx 157.9$ MPa.
    The orange softening curve has $H_d = 105$, resulting in $g_1 \approx 142.9$ MPa.}
	\label{fig:hardening}
\end{figure}

Four material parameters were initially selected for the calibration study: the inverse rate sensitivity, $m$; the initial CRSS, $g_0$; and the hardening and dynamic recovery coefficients, $H$ and $H_d$, respectively.
However, the hardening model has a saturation stress, $g_1$, that is found by setting the hardening rate to zero in Eq.~\eqref{eq:hardening} and solving for $g_s$:
\begin{equation}
    \label{eq:saturation}
    g_1 = H / H_d.
\end{equation}
Strain hardening and softening correspond to $g_0 < g_1$ and $g_0 > g_1$, respectively, as visualized in \autoref{fig:hardening}.
Sampling independently from prior distributions on $H$ and $H_d$ may lead to cases where $g_0 \gg g_1$, causing extreme softening immediately after the yield point.
This behavior is nonphysical for the material of interest and causes numerical problems in the solver.
Therefore, the ratio $g_0 / g_1$ was used as a calibration parameter instead of $H_d$, providing three intuitive characteristics:
\begin{enumerate}
	\item Strain hardening occurs if $g_0 / g_1 < 1$. The amount of hardening increases as the ratio approaches zero.
	\item Strain softening occurs if $g_0 / g_1 > 1$. The amount of softening increases as the ratio increases.
	\item Perfectly plastic behavior occurs if $g_0 / g_1 = 1$.
\end{enumerate}
An upper bound on the support of the prior distribution for $g_0 / g_1$ restricted the maximum amount of softening.
%

Following standard practice for calibration experiments where the noise level is unknown, $\sigma_\text{noise}$ is included in the set of calibration parameters.
The vector of calibration parameters was therefore
\begin{equation}
\bm{\theta} = \left[m, g_0, H, g_0 / g_1, \sigma_\text{noise}\right],
\end{equation}
so the number of parameters was $p = 5$.
When data is available from multiple experiments as is the case for the multiscale calibration presented here, separate noise levels for each measurement type can be estimated as appropriate.
An example of this case is analyzed in the \nameref{sec:two_stage_realistic} section.

Independent uniform priors were placed on each parameter based on the ranges considered in \cite{bandyopadhyay_uncertainty_2019}.
This means that, prior to observing data, it is assumed that all values within the prior bounds are equally likely to be the true value.
Prior bounds are shown in \autoref{tab:priors}.
%
\begin{table}
	\centering
	\caption{Ground-truth values and bounds for the uniform priors on the calibration parameters.}
		\begin{tabular}{cccc}
		\toprule
		Parameter & Lower bound & Upper bound & Ground truth \\
		\midrule
		$m$ & $5$ & $25$ & $10$ \\
		$g_0$ & $300$ MPa & $700$ MPa & $400$ MPa \\
		$H$ & $1000$ MPa & $30,000$ MPa & $10,000$ MPa \\
		$g_0 / g_1$ & $0.0$ & $1.5$ & $0.8$ \\
		$\sigma_\text{noise}$ & $0$ MPa & $100$ MPa & $5$ MPa \\
		\bottomrule
	\end{tabular}
	\label{tab:priors}
\end{table}

\subsubsection*{Calibration Data}
\label{sec:calibration_data}
%
\begin{figure}
	\centering
	\includegraphics[width=0.7\textwidth]{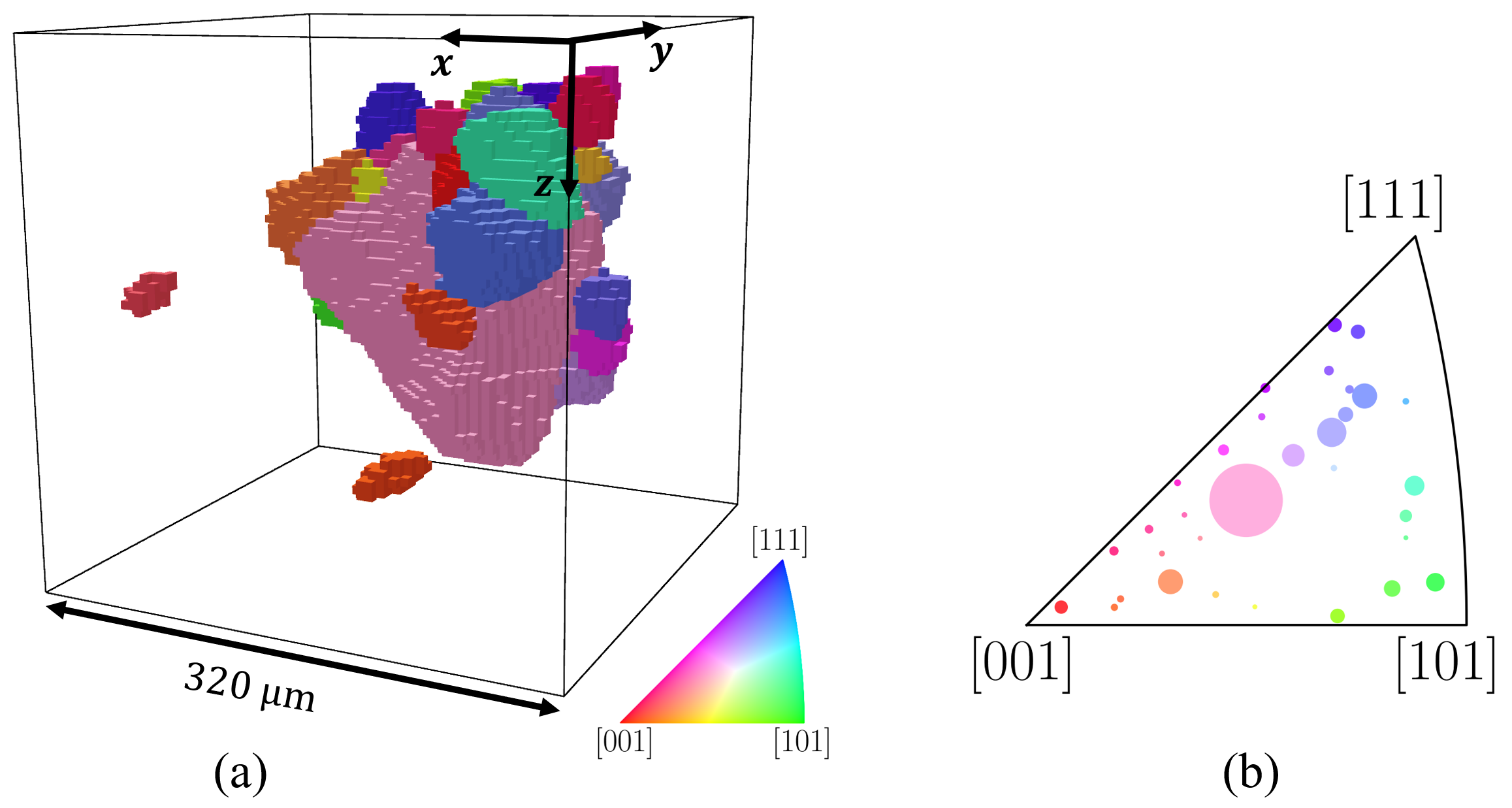}
	\caption{The 32 grains whose grain-average stresses were used as local data in the calibration.
	(a) IPF map with a bounding box representing the full microstructural domain.
	(b) IPF with points scaled by grain size and colored according to the IPF map.
	The IPF map and IPF are both with respect to the loading ($x$) direction.
	}
	\label{fig:grains_of_interest}
\end{figure}
Synthetic measurement data was generated by a single CP simulation with fixed ground-truth parameter values, again based on \cite{bandyopadhyay_uncertainty_2019} and provided in \autoref{tab:priors}.
The global data consists of the mean stress in the loading direction at 13 time points during the simulation corresponding to mean loading direction strains of $0.3\%$, $0.4\%$, $0.425\%$, $0.45\%$, $0.475\%$, $0.5\%$, $0.55\%$, $0.6\%$, $0.65\%$, $0.7\%$, $0.8\%$, $0.9\%$, and $1.0\%$.
The strains were chosen to have a higher density of data near the onset of global yielding.
The local data consists of grain-average stresses in the loading direction at the same 13 time points for 32 grains, resulting in 416 total local data points.
The 32 grains, shown in \autoref{fig:grains_of_interest}, are all at least 5 voxels from any edge of the microstructural region and have a volume of at least 20 voxels.
These selection criteria were intended to mimic grain selection in HEDM experiments \cite{menasche_afrl_2021}.
Finally, the mean and grain-average stresses were polluted with zero-mean noise sampled from a Normal distribution with standard deviation $\sigma_\text{noise} = 5$ MPa.
The resulting dataset was used as the measurement data for all calibrations, with some modifications to the noise and availability of the local data in the \nameref{sec:two_stage_realistic} section.

\begin{figure}
	\centering
	\includegraphics[width=0.95\textwidth]{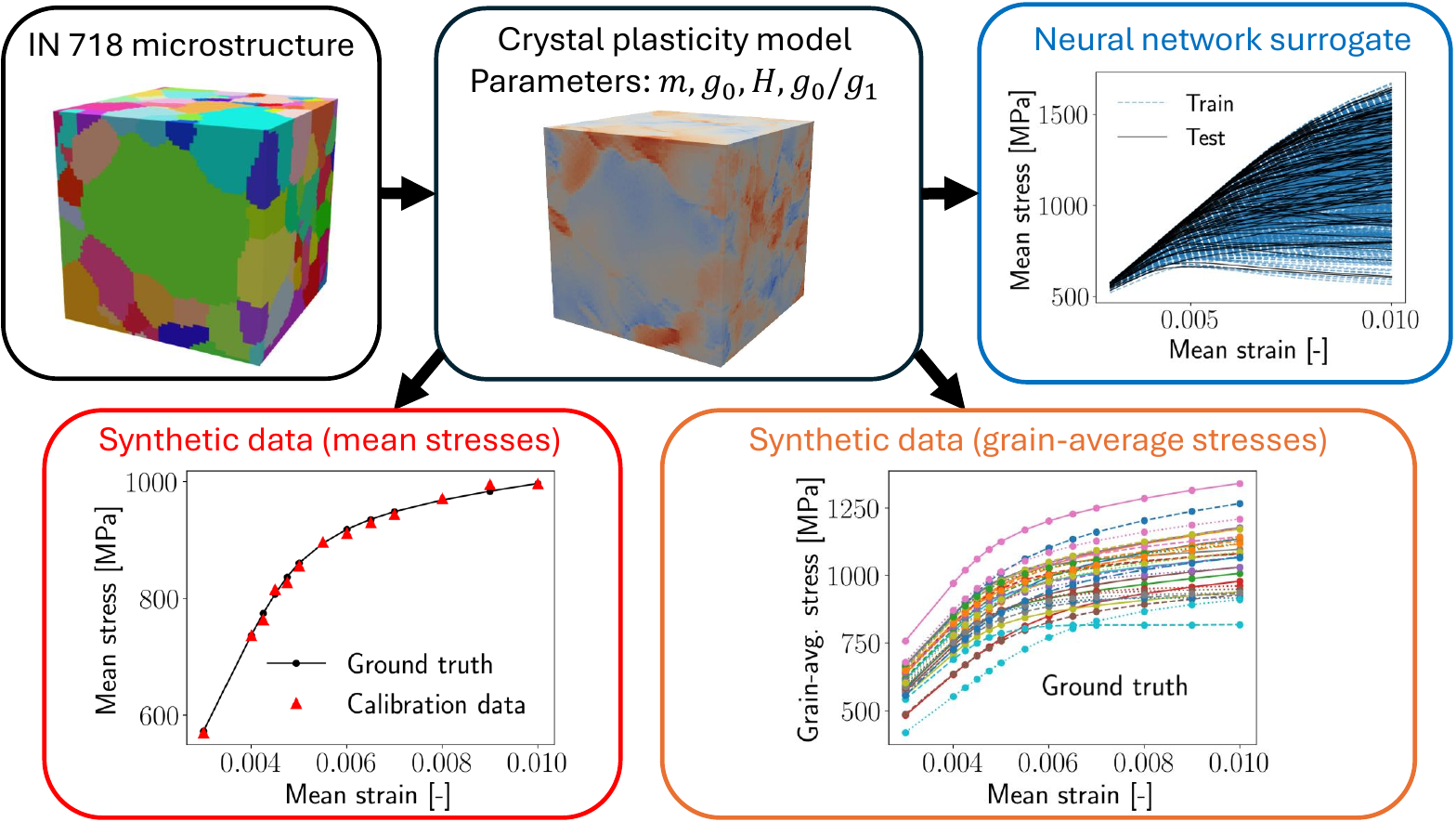}
	\caption{Summary of the CP model setup (black boxes) and data generated for use in the calibration.
	The training and test data for the surrogate model (blue box) consists of the mean stresses from 1000 simulations with material parameters sampled from the uniform prior distributions.
    The synthetic measurement data is from a simulation with $m = 10$, $g_0 = 400$ MPa, $H = 10,000$ MPa, and $g_0 / g_1 = 0.8$, with zero-mean noise added to both the mean stresses (red box) and the grain-average stresses (orange box).
	The mean stresses are shown before and after adding the zero-mean noise.
	For clarity, the grain-average stresses are shown prior to adding the noise.
	}
	\label{fig:synthetic_data}
\end{figure}

The synthetic data generated from the CP model, including the calibration data and the training data for the surrogate model, is summarized in \autoref{fig:synthetic_data}.
For the mean stresses, the ground-truth data (before adding the noise) and the calibration data (after adding the noise) are both shown.
For clarity, only the ground-truth grain-average stresses are shown.
The surrogate model training is described further in the \nameref{sec:sensitivity} section.%
%

\subsubsection*{Two-Stage Calibration Approach}
\label{sec:calibration_two_stage}
A goal of the two-stage calibration in this study is to increase computational efficiency of SMC with the full-field CP model relative to a single-stage calibration.
The computational cost of the calibration is a function of the number of likelihood evaluations.
The number of likelihood evaluations in turn depends on the number of SMC steps required for convergence of the algorithm.
More SMC steps are typically required if the prior distribution is uninformative or minimally informative, meaning that it has a much wider distribution of probability mass than the posterior distribution.
Therefore, using informative priors that place the initial particle population closer to the posterior could reduce the computational cost of SMC.

The goal of stage one was to infer a set of preliminary posterior distributions that are in turn used as informative priors for stage two.
Stage one was computationally cheap due to using the neural network surrogate model, which has negligible cost to evaluate relative to the full-field CP model.
On the other hand, there is a risk that the informative priors will be overconfident and biased (i.e., tightly distributed about incorrect parameter values) due to bias in the surrogate model.
To help avoid this issue, only the global data was used in stage one: the training data for the neural network surrogate came from the mean stresses in a set of CP simulations, and only the mean stresses were provided as measurement data in the SMC calibration.
Thus, the number of measurement points for stage one was $q = 13$.

A kernel density estimate (KDE) of the joint posterior distribution of the material parameters from stage one was used as the prior for the material parameters in stage two.
Since one calibration case considered different noise for the mean strains and grain-average strains, the original uniform prior was still used for the noise standard deviation in stage two to enable fair comparisons between each calibration.
The full synthetic dataset of mean and grain-average stresses was used as measurement data for stage two, resulting in $q = 429$ measurement points.
The computational efficiency of the two-stage procedure is quantified by comparing walltime and the number of SMC steps with one-stage calibrations in the \nameref{sec:one_stage} section.
Refer to \autoref{fig:figure1} for a visual summary of the overall two-stage calibration procedure.

\section*{Results}
\label{sec:results}
Results are first presented for the surrogate model for the mean stresses, including trends in the training and test samples from the CP model, to gain insight into possible challenges in the calibration procedure.
Then, posterior distributions are presented for each stage in the two-stage calibration to study the influence of local data on the inferred parameter values and associated uncertainty.
Stage two in the calibration is repeated with different restrictions on the local data to examine the robustness of the calibration procedure to more realistic data.
Finally, the two-stage procedure is compared with several one-stage approaches to study tradeoffs for computational cost and calibration accuracy and precision.

Throughout this section, the marginal and joint posterior distributions from each calibration are primarily visualized and compared using histograms.
Summary statistics for all the marginal posterior distributions are also provided in Appendix \ref{app:statistics} to assist with more quantitative comparisons.

\subsubsection*{Sensitivity Study and Surrogate Model Training}
\label{sec:sensitivity}
For the two-stage calibration procedure, a neural network surrogate model (specifically, the multilayer perceptron in \verb|scikit-learn| \cite{pedregosa_scikit-learn_2011}) was trained to map the four calibration parameters to the 13 global mean stresses.
To generate the training data, CP simulations were completed using 1000 Latin hypercube samples within the bounds of the independent uniform prior distributions on the four material parameters (see \nameref{sec:calibration_parameters}).
The data was split into a set of 900 training samples and 100 test samples.
Trends in the training and test data are analyzed first.
%
\begin{figure}
	\centering
	\includegraphics[width=0.5\textwidth]{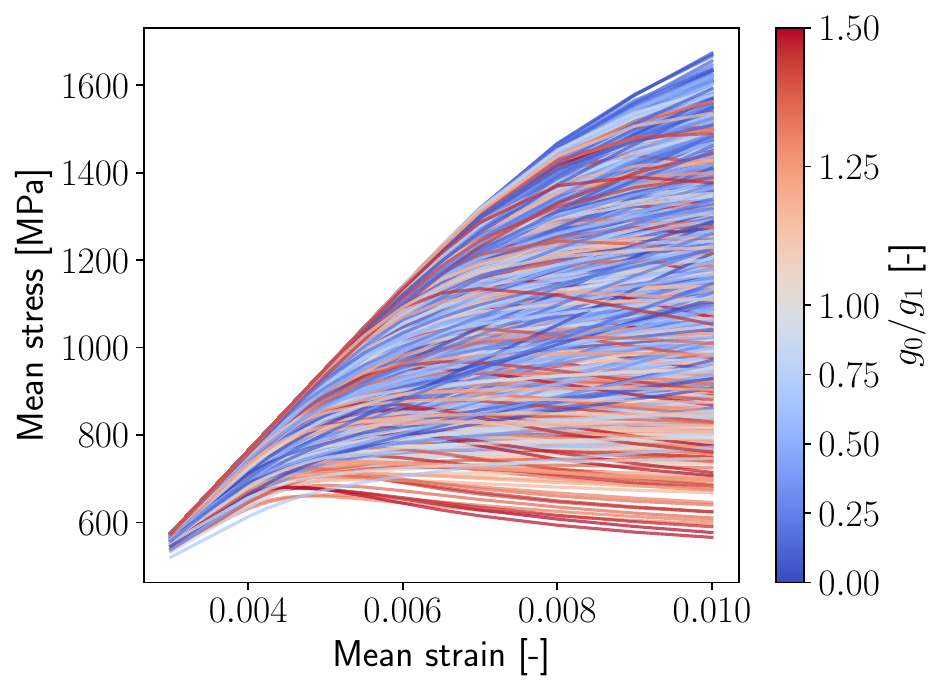}
	\caption{Mean stress-strain curves at the 13 time points of interest for the 1000 simulations from Latin hypercube samples of the four material parameters used to train and test the neural network surrogate model.
	The curves are colored such that red indicates $g_0 / g_1 > 1$ and blue indicates $g_0 / g_1 < 1$ to highlight the different curve shapes.}
	\label{fig:training}
\end{figure}

Mean stress-strain curves at the 13 time points of interest for the 1000 Latin hypercube samples are shown in \autoref{fig:training}.
At lower strains, all simulations are in the elastic regime or in the initial stages of global yielding.
The curves start to diverge from one another as plasticity sets in at higher strain values.
Samples with softening ($g_0 / g_1 > 1.0$) have a downward slope after yielding begins and are particularly noticeable for low stress values at $1.0\%$ strain.

\begin{figure}
	\centering
	\includegraphics[width=\textwidth]{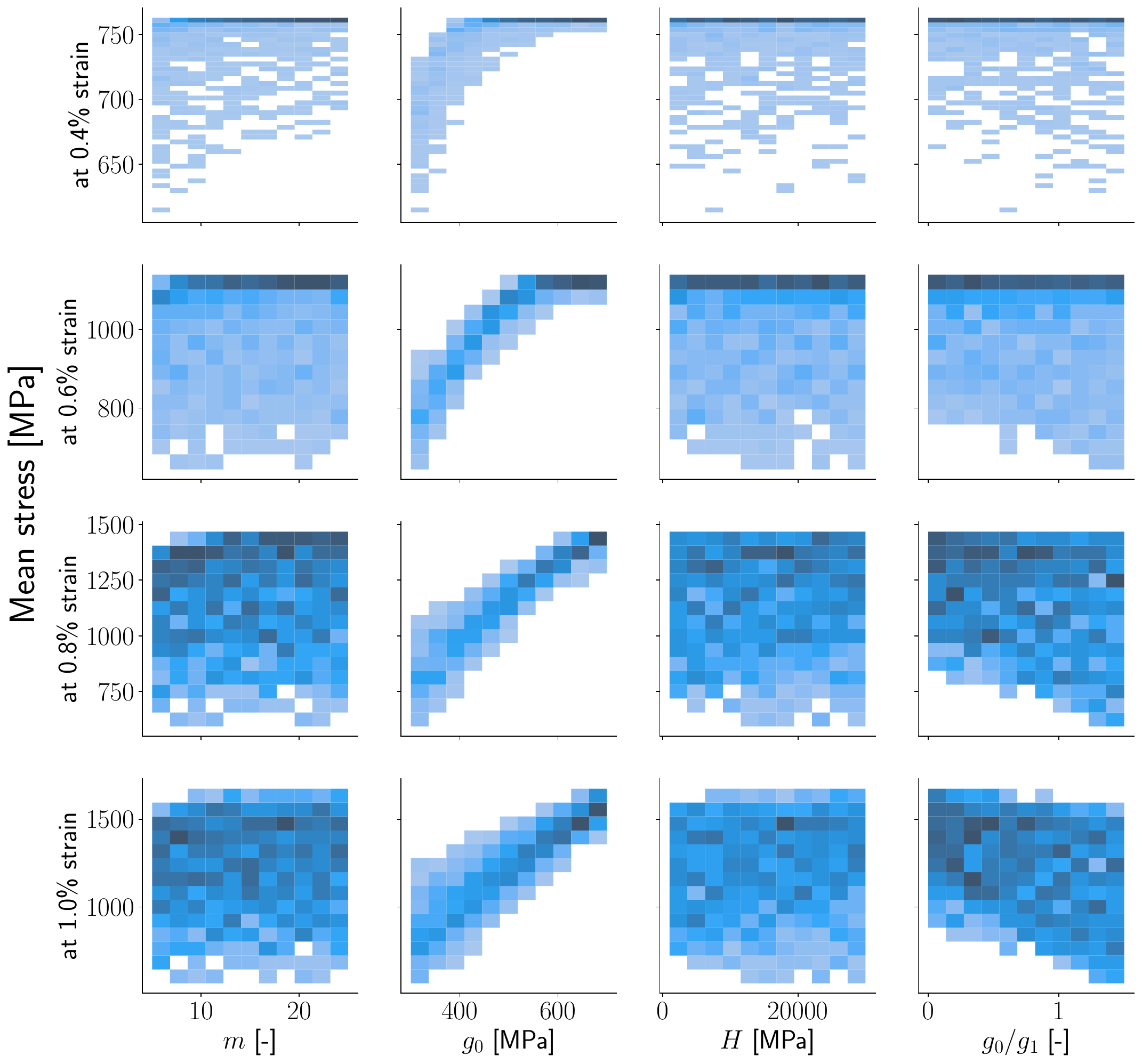}
	\caption{Pairwise plots of the material parameters and the mean stresses at mean strains of $0.4\%$, $0.6\%$, $0.8\%$, and $1.0\%$ in the 1000 Latin hypercube samples.
    The stresses are strongly correlated with $g_0$, while first-order correlations with other parameters appear to be weak.}
	\label{fig:pairwise_stress_params}
\end{figure}

To further analyze the relationship between the mean stresses and the calibration parameters, the mean stresses at $0.4\%$, $0.6\%$, $0.8\%$, and $1.0\%$ mean strain are extracted for each simulation.
The pairwise plots in \autoref{fig:pairwise_stress_params} show correlations between these four mean stresses and the material parameters.
The main finding from the pairwise plots is that all stresses are strongly correlated with the initial CRSS.
Other first-order correlations are relatively weak.
By controlling the initial yield point, the initial CRSS essentially scales the stress-strain curves in the plastic regime up or down.
The ratio $g_0 / g_1$ is related to the strength of the dynamic recovery term and is weakly correlated with the stress at $1.0\%$ strain.
Meanwhile, a lower value of the inverse rate sensitivity allows yielding to begin at lower stresses, shown through the correlation between $m$ and the stress at $0.4\%$ strain.
Overall, however, the lack of clear correlations between the mean stresses and parameters other than the initial CRSS suggest that uniqueness challenges may arise in the calibration.%
%

With the training data in hand, the surrogate was trained using the Adam optimizer and an adaptive learning rate with rectified linear unit (RELU) activation functions.
To minimize effects of feature scaling, the training data was scaled to zero mean and unit variance.
Optimal hyperparameters of the surrogate were found using a grid search with five-fold cross-validation on the training set.
The hyperparameters included in the grid search were the number and size of the hidden layers, the strength of the L2 regularization term in the loss function ($\alpha$), and the initial learning rate ($\eta$).
Values of each parameter included in the grid search are based on the \verb|scikit-learn| documentation and examples and are given in \autoref{tab:grid_search}.
Using the optimal hyperparameters, the model was then trained on the entire training set and evaluated on the test set.
%
\begin{table}
	\centering
	\caption{Grid search values for parameters in the neural network surrogate model, with optimal values bolded.}
		\begin{tabular}{ccc}
		\toprule
        Parameter & Grid search values \\
		\midrule
		Hidden layer sizes & $\left[(50,), (100,), (50, 50), (100, 50), \mbf{(100, 100)}\right]$ \\
		$\alpha$ & $\left[\mbf{10^{-1}}, 10^{-2}, 10^{-3}, \ldots, 10^{-7} \right]$ \\
        $\eta$ & $\left[ 10^{-1}, 10^{-2}, \mbf{10^{-3}}, 10^{-4} \right]$ \\
        \bottomrule 
	\end{tabular}
	\label{tab:grid_search}
\end{table}

Due to the scaling effect of the initial CRSS, the stresses at each strain value are correlated with each other.
However, the neural network surrogate assumes that all output quantities are independent.
Thus, the surrogate was trained using the first four principal components (PCs) of the stresses from the training data, determined using principal component analysis (PCA).
The first four PCs explain more than $99.9\%$ of the variance in the mean stress data.
Using the PCs was also found to eliminate convergence issues for some parameter combinations in the grid search and to produce stress-strain curves with more realistic shapes.
Details on the PCA can be found in Appendix \ref{app:pca}.
Using the optimal hyperparameters from \autoref{tab:grid_search}, the root-mean-squared error was 5.42 MPa on the training set and 5.37 MPa on the test set (about $0.6\%$ of the ground-truth global yield stress of approximately 900 MPa), suggesting that overfitting was avoided.

\subsubsection*{Two-Stage Calibration}
\label{sec:two_stage}
Stage one required 14 SMC steps and about 10 seconds on a laptop computer to generate 40,000 posterior samples using the surrogate model.
Stage two was completed on the NASA Langley K-cluster, using 644 CPU cores across 23 Intel E5-2697 V3 nodes.
Each core was assigned two particles in the SMC algorithm, resulting in 1288 total posterior samples.
As noted in the \nameref{sec:smc} section, one SMC step required five model evaluations per particle and therefore ten total evaluations on each core.
Stage two required 18 SMC steps and $115,920$ CP simulations, with a total walltime of about 90 hours and total CPU time of about 2400 days.
This computation would be intractable without the parallelization enabled by the SMC algorithm.

\begin{figure}
	\centering
	\begin{subfigure}{\textwidth}
		\includegraphics[width=\textwidth]{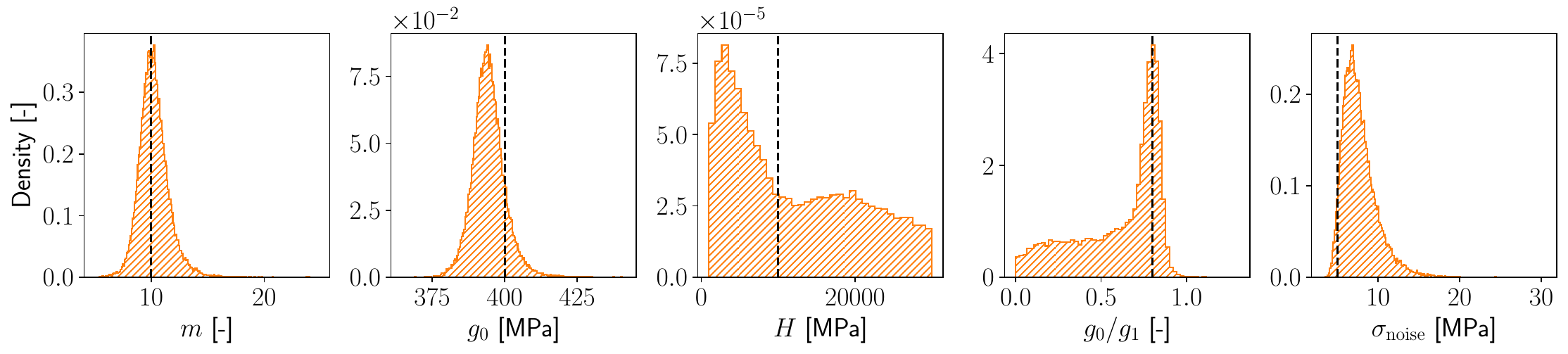}
		\caption{Stage one (surrogate with mean stresses only).}
	\end{subfigure}
	\begin{subfigure}{\textwidth}
		\includegraphics[width=\textwidth]{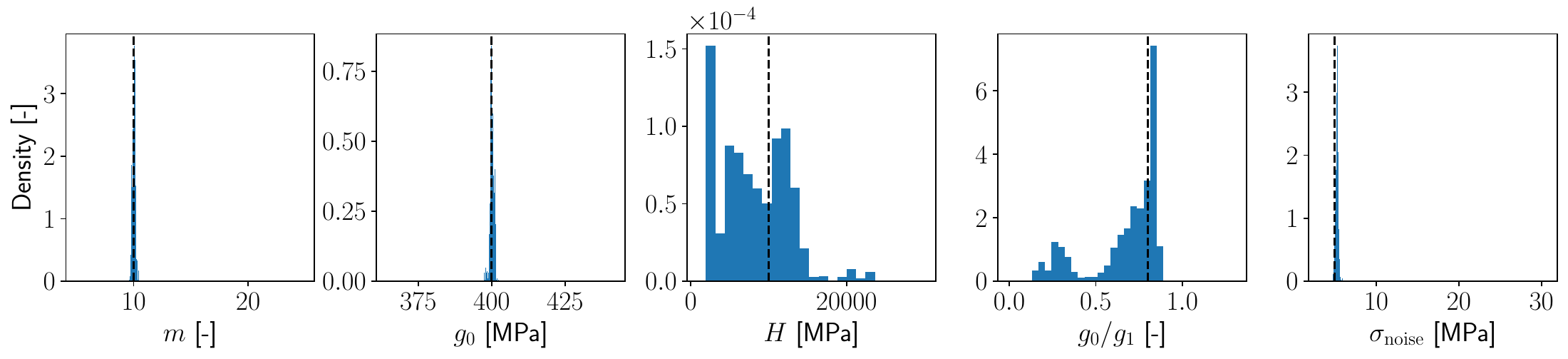}
		\caption{Stage two (CP model with mean and grain-average stresses).}
	\end{subfigure}
    \begin{subfigure}{\textwidth}
		\includegraphics[width=\textwidth]{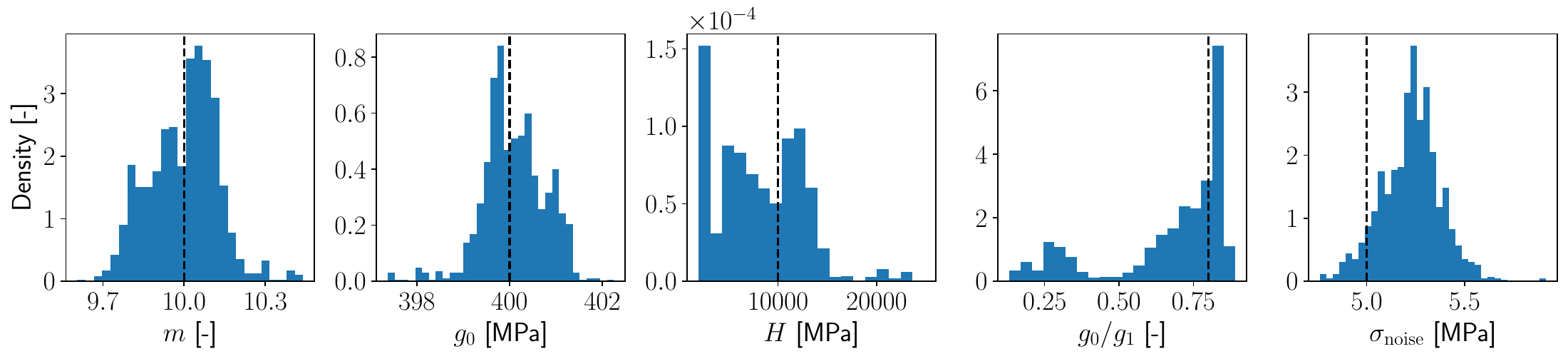}
		\caption{Stage two, with tighter horizontal axis limits to more clearly show the distributions.}
	\end{subfigure}
	\caption{Marginal posterior distributions for the five calibration parameters from each stage in the two-stage calibration.
    The dashed lines represent the ground-truth values of the parameters.}
    \label{fig:two_stage_marginals}
\end{figure}

Marginal posterior distributions from both stages are shown in \autoref{fig:two_stage_marginals}.
In stage one (\autoref{fig:two_stage_marginals}a), $m$ and $g_0$ appear to be reasonably well calibrated.
Both have approximately Normal distributions centered close to the ground-truth value.
However, there is some bias in the posterior distribution for $g_0$, and some samples of $m$ are close to the upper bound of the prior and far from the ground-truth value.
On the other hand, calibration of $H$ and $g_0 / g_1$ appears to be more challenging.
Although the mode of the distribution for $g_0 / g_1$ is close to the ground-truth value, the posterior distribution for $H$ is biased toward lower values, and the posterior distributions for both parameters span nearly the entire parameter space.
Samples with slip-system-level softening behavior ($g_0 / g_1 > 1.0$) are mostly, but not entirely, eliminated.
Finally, the posterior distribution for the scale of the noise, $\sigma_\text{noise}$, is biased above the ground-truth value of 5 MPa, and the upper tail extends to significantly larger values.
This bias indicates that the noise term is compensating for discrepancy between the model and the data-generating process.%
%

Posterior distributions after stage two (\autoref{fig:two_stage_marginals}b and c) show significantly reduced uncertainty for $m$ and $g_0$, forming a tighter distribution centered on the ground-truth values relative to stage one.
Uncertainty is also somewhat reduced for $H$ and $g_0 / g_1$.
In particular, samples with softening are completely eliminated: the maximum value of $g_0 / g_1$ across the 1288 posterior samples is about 0.89.
However, the distributions for $H$ and $g_0 / g_1$ both appear to be bimodal, and the posterior samples include values of $H$ from across almost the entire parameter space.
This suggests that obtaining point estimates of $H$ and $g_0 / g_1$ remains challenging, even with both global and local calibration data.
Finally, the bias and uncertainty in the posterior distribution for $\sigma_\text{noise}$ are both significantly reduced, reflecting the use of the full CP model rather than the biased surrogate.

\begin{figure}
	\centering
	\begin{subfigure}{0.49\textwidth}
		\includegraphics[width=\textwidth]{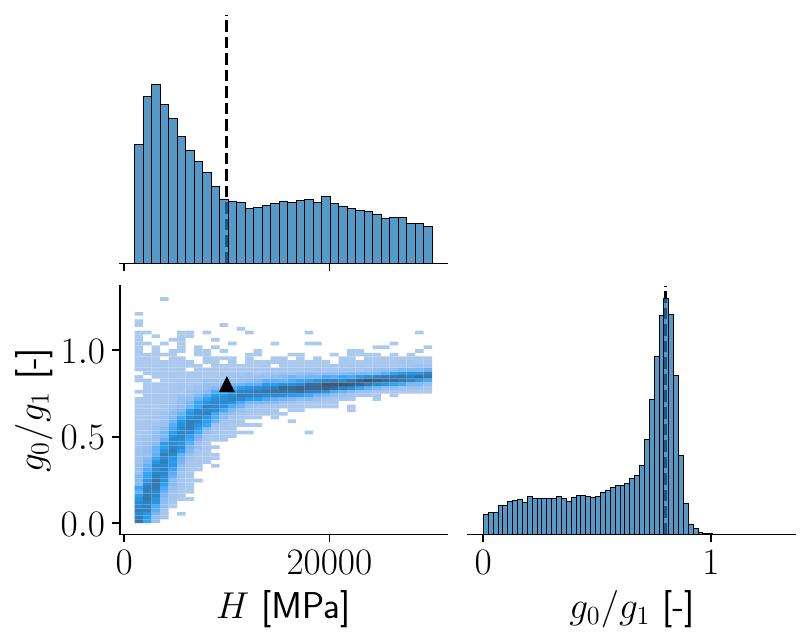}
		\caption{Stage one.}
	\end{subfigure}
	\hfill
	\begin{subfigure}{0.49\textwidth}
		\includegraphics[width=\textwidth]{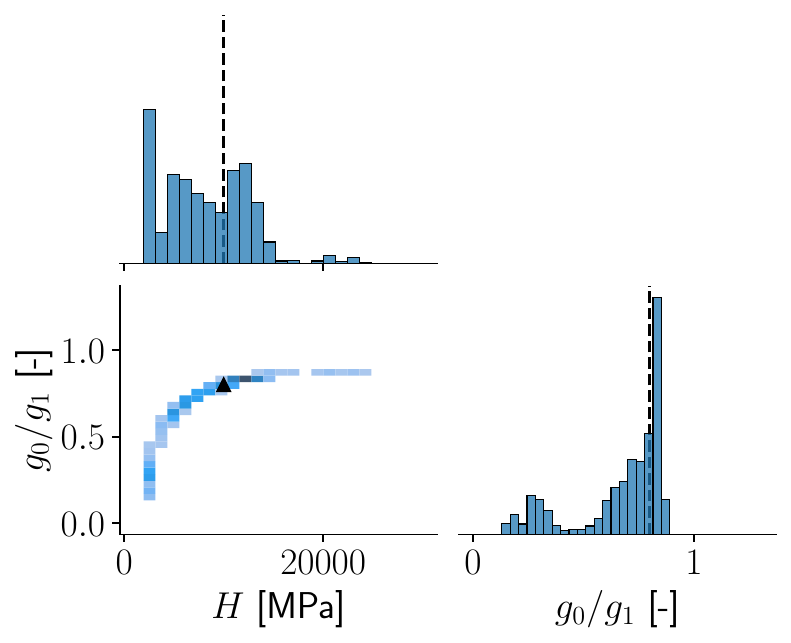}
		\caption{Stage two.}
	\end{subfigure}
	\caption{Pairwise plots for the posterior distributions of $H$ and $g_0 / g_1$ after the first and second calibration stages.
    Black dashed lines in the marginal distributions along the diagonals and black triangles in the joint distributions represent the ground-truth values.}
	\label{fig:pairwise_H_ratio}
\end{figure}

To obtain further insight into the posterior distributions of $H$ and $g_0 / g_1$, pairwise plots for these parameters after stage one and stage two are shown in \autoref{fig:pairwise_H_ratio}a and \autoref{fig:pairwise_H_ratio}b, respectively.
The off-diagonal joint distributions demonstrate an interesting correlation structure, which becomes stronger after stage two.
The joint distribution in \autoref{fig:pairwise_H_ratio}b has one mode close to the ground-truth values of both parameters.
However, the shape of the distribution suggests that a range of $H$ values is consistent with the data for $g_0 / g_1 \approx 0.8$.
There is also a second mode of the distribution at small values of $H$, where a range of $g_0 / g_1$ values appears to be consistent with the data.
Finally, the sharp drop-off in the distribution as $g_0 / g_1$ approaches 0.9 suggests that the calibration may have located an upper bound for this parameter below 1.0, due in part to the minimal uncertainty about $g_0$.
Essentially, a too-high value of $g_0 / g_1$ would cause the grain-average stresses to saturate before reaching stress values seen in the data, regardless of the value of $H$.
%

\subsubsection*{Two-Stage Calibration with More Realistic Grain-Average Stress Data}
\label{sec:two_stage_realistic}
The results in the previous section represent an ideal situation where grain-average stress data is available at the same time points and the same noise level as the global mean stress.
In reality, estimating grain-average stresses from HEDM measurements is significantly more expensive than measuring the mean stress with a load cell and requires pausing the test at each point of interest.
The HEDM data also may be influenced by more sources of noise and uncertainty than the mean stress measurement.
Therefore, stage two of the calibration was repeated twice to study the influence of more realistic data constraints.

First, calibration data for the grain-average stresses was made available only for mean strains of $0.4\%$, $0.6\%$, $0.8\%$, and $1.0\%$ (i.e., there are only 4 grain-average stress values for each of the 32 grains of interest, resulting in 128 total local data points).
As a result, there were $q = 141$ total measurement points.
The Air Force Research Laboratory additive manufacturing modeling challenge had similar constraints on the amount of HEDM grain-average stress data collected in the plastic regime \cite{menasche_afrl_2021}.
This case is referred to as the ``reduced grain stresses'' case.

Second, the standard deviation of the noise added to the grain-average stress measurements was doubled to 10 MPa.
This case is referred to as the ``increased noise'' case.
The variance of the noise term in the likelihood function, Eq.~\eqref{eq:likelihood}, was modified such that $\Sigma_{1:13,1:13} = \bm{I}_{13 \times 13}\sigma_\text{noise,global}^2$ for the 13 mean stress data points and $\Sigma_{14:429,14:429} = \bm{I}_{416 \times 416}\sigma_\text{noise,local}^2$ for the 416 grain-average stress data points.
Both $\sigma_\text{noise,global}$ and $\sigma_\text{noise,local}$ were included as calibration parameters with the same uniform priors and ground-truth values of $\sigma_\text{noise,global} = 5$ MPa and $\sigma_\text{noise,local} = 10$ MPa.
Thus, the number of calibration parameters was $p = 6$, and the number of measurement points was $q = 429$.
%
\begin{figure}
	\centering
	\begin{subfigure}{\textwidth}
		\includegraphics[width=\textwidth]{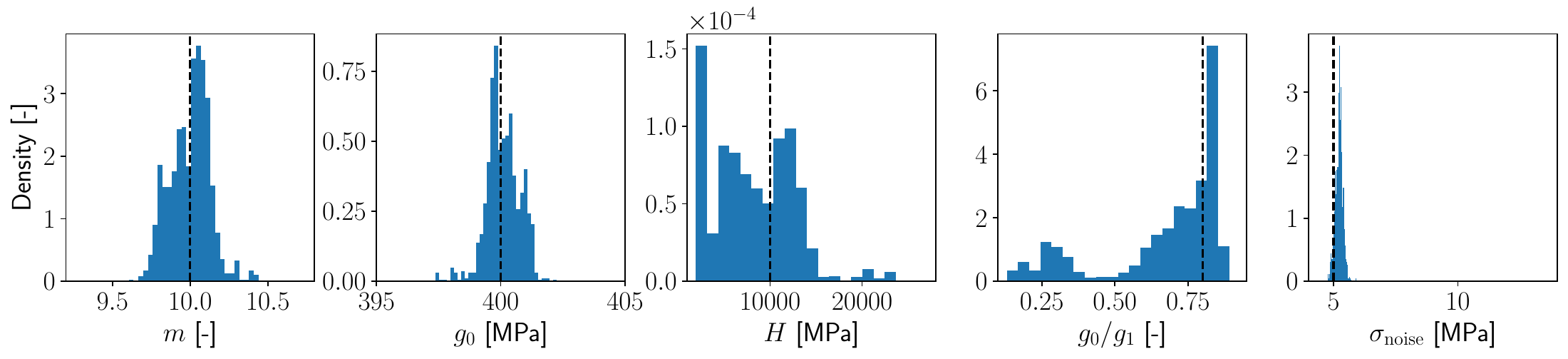}
		\caption{Two-stage calibration with all stress data and equivalent global and local noise (same data as in \autoref{fig:two_stage_marginals}b and \autoref{fig:two_stage_marginals}c).}
	\end{subfigure}
	\begin{subfigure}{\textwidth}
		\includegraphics[width=\textwidth]{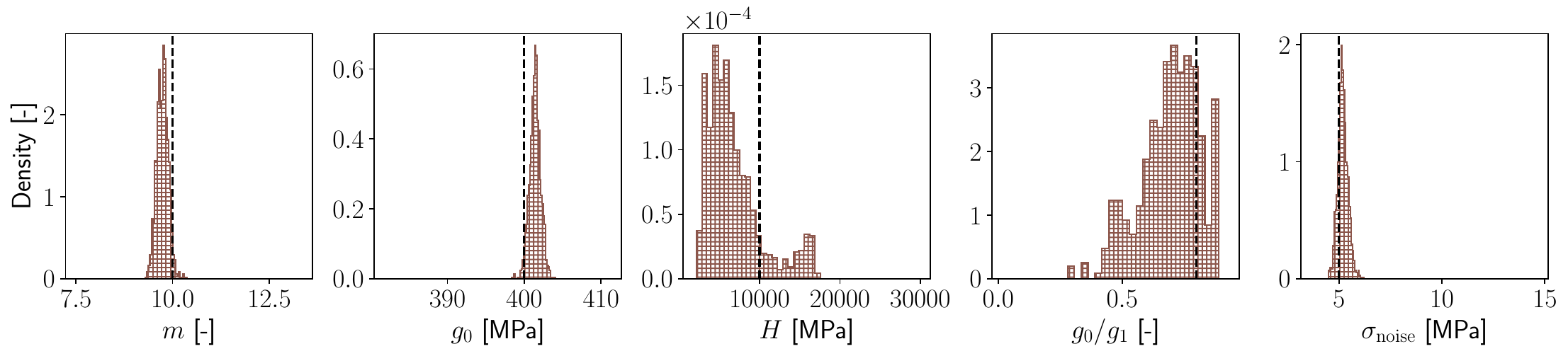}
		\caption{Reduced grain stresses case.}
	\end{subfigure}
    \begin{subfigure}{\textwidth}
		\includegraphics[width=\textwidth]{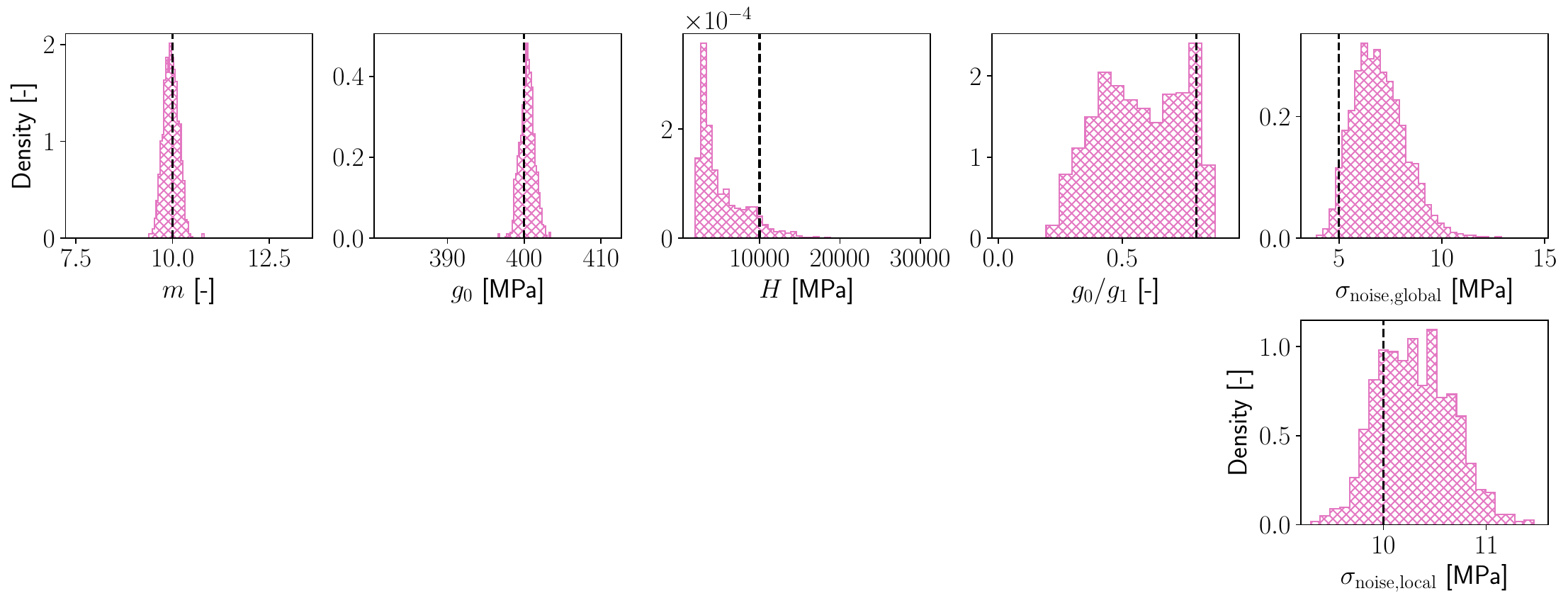}
		\caption{Increased noise case.}
	\end{subfigure}
	\caption{Marginal posterior distributions for the five calibration parameters from the two-stage calibration with different constraints on the grain-average stress data.
	The dashed lines represent the ground-truth values of the parameters.}
    \label{fig:marginals_comparison}
\end{figure}

\autoref{fig:marginals_comparison} shows a comparison between the marginal posterior distributions for the two new cases stage two results.
For $m$ and $g_0$, the calibration procedure is fairly robust to reduced data and increased noise.
The uncertainty is similar across all three cases, although some bias occurs in the reduced grain stresses case.
The increased noise case does not have this bias, suggesting that the quantity of local measurements may be more important than their precision.
The posterior distributions for $H$ and $g_0 / g_1$ in the two new cases still eliminate samples at large values of both parameters.
However, the posterior distributions for $H$ in particular have notable bias towards values below the ground truth.
The strong correlation between $H$ and $g_0 / g_1$ observed in \autoref{fig:pairwise_H_ratio}b is also still present in the joint posterior distribution.
As before, the ground-truth value is in the support of the joint posterior distribution for both the reduced grain stresses and increased noise cases, even though point estimates from the marginal distributions of $H$ and $g_0 / g_1$ would be biased.

\FloatBarrier
\subsubsection*{Alternative One-Stage Calibration Approaches}
\label{sec:one_stage}
\begin{figure}
	\centering
	\begin{subfigure}{\textwidth}
		\includegraphics[width=\textwidth]{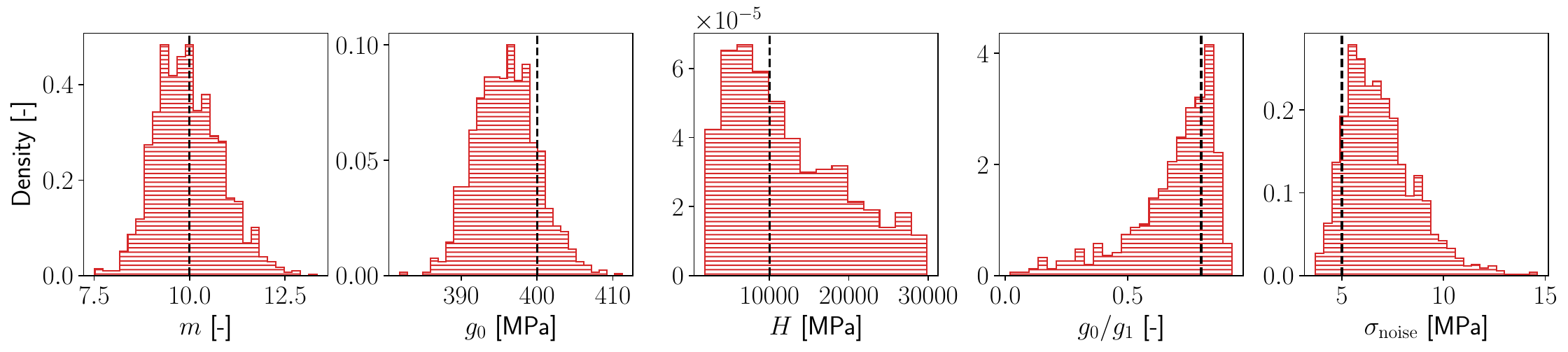}
		\caption{CP model with only mean stresses (CP-Mean).}
	\end{subfigure}
	\begin{subfigure}{\textwidth}
		\includegraphics[width=\textwidth]{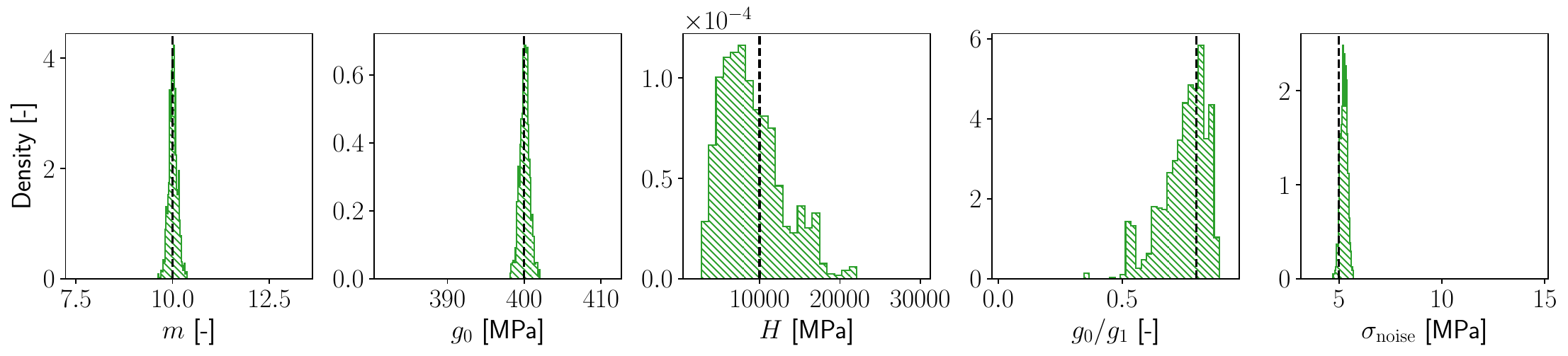}
		\caption{CP model with mean and grain-average stresses (CP-All).}
	\end{subfigure}
    \begin{subfigure}{\textwidth}
		\includegraphics[width=\textwidth]{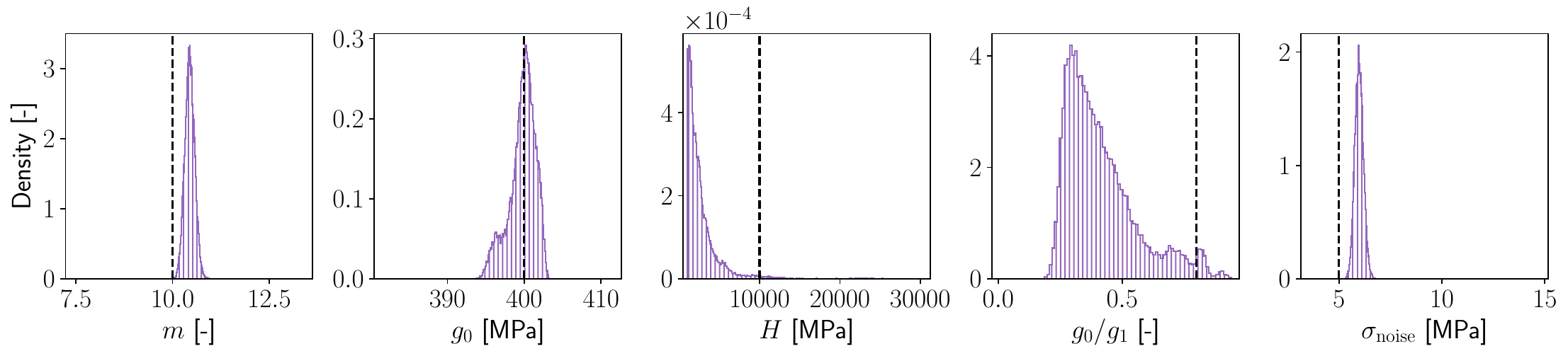}
		\caption{Surrogate model with mean and grain-average stresses (Surrogate-All).}
	\end{subfigure}
	\begin{subfigure}{\textwidth}
		\includegraphics[width=\textwidth]{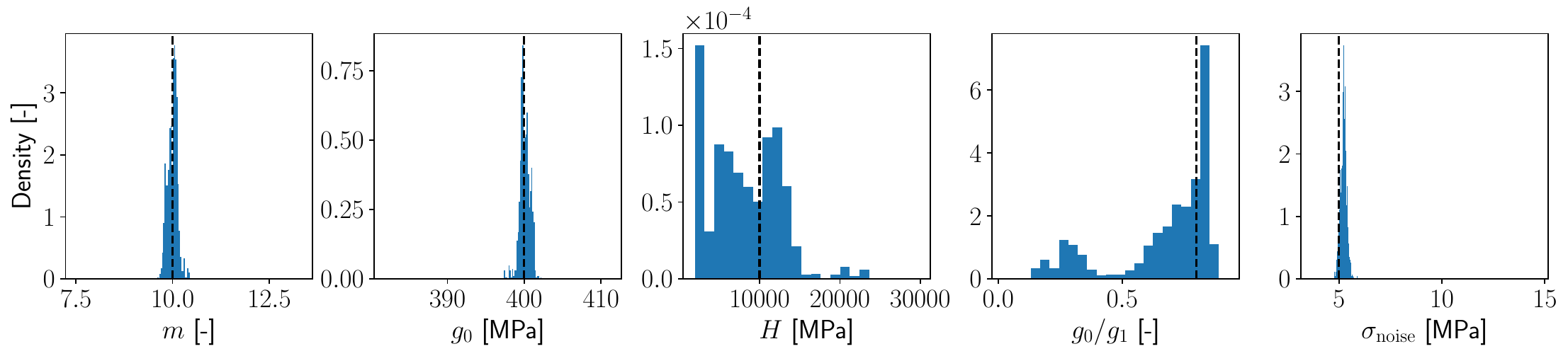}
		\caption{Stage two of the two-stage calibration (same data as in \autoref{fig:two_stage_marginals}bc).}
	\end{subfigure}
	\caption{Marginal posterior distributions for the five calibration parameters from the three one-stage calibration procedures described in the \nameref{sec:one_stage} section.
    All cases use the same horizontal axis bounds.
	A comparison with the two-stage calibration is also included in (d).
	The dashed lines represent the ground-truth values of the parameters.}
    \label{fig:marginals_one_stage}
\end{figure}

To assess the advantages and limitations of the proposed two-stage calibration, it is interesting to compare the results with several one-stage calibrations in the \nameref{sec:one_stage} section.
To this end, three one-stage calibrations, using the uniform priors in \autoref{tab:priors}, are presented here:
\begin{enumerate}
    \item ``CP-Mean'': Calibration using the CP model but with only the mean stresses as calibration data (identical to stage one in the two-stage calibration but with the CP model instead of the surrogate).
    \item ``CP-All'': Calibration using the CP model with both mean and grain-average stresses (identical to stage two in the two-stage calibration but with uniform priors).
    \item ``Surrogate-All'': Calibration using a surrogate trained on the first six PCs of the full dataset of both mean and grain-average stresses from the Latin hypercube samples. The training procedure was identical to the one described in the \nameref{sec:sensitivity} section.
\end{enumerate}
Marginal posterior distributions from the three one-stage approaches are shown in \autoref{fig:marginals_one_stage}.
To help with comparisons, the marginal posterior distributions from the original two-stage calibration are shown again in \autoref{fig:two_stage_marginals}d.
%

The posterior distributions for CP-Mean (\autoref{fig:marginals_one_stage}a) demonstrate the limitations of using only global data.
There is more uncertainty in the distributions for $m$ and $g_0$ than in any of the two-step calibration results.
No samples have $g_0 / g_1 > 1$, but the distributions for both $H$ and $g_0 / g_1$ still span most of the parameter space.

The posterior distributions for CP-All (\autoref{fig:marginals_one_stage}b) are very similar to the end of the two-stage calibration.
Both approaches should converge to the same result in the limit of a large amount of calibration data since the only difference between the two calibrations is the priors: uniform priors for \autoref{fig:marginals_one_stage}b and posteriors from stage one for \autoref{fig:marginals_one_stage}d.
The most significant difference in the posterior distributions is that the second mode at small values of $g_0 / g_1$ and $H$ is largely absent in \autoref{fig:marginals_one_stage}b.
The CP-All case required 22 SMC steps and about $7.5$ days of walltime (about 4800 days of CPU time), compared to 18 SMC steps and about $3.75$ days of walltime (about 2400 days of CPU time) for the two-stage calibration using the same computing resources.

Finally, \autoref{fig:marginals_one_stage}c shows the posterior distributions for Surrogate-All.
The surrogate model was trained using the same strategy as in the \nameref{sec:sensitivity} section; here, the first six PCs explained greater than $99.9\%$ of the variance.
The root-mean-squared error was $5.94$ MPa on the training data and $5.96$ MPa on the test data (about $0.7\%$ of the approximate ground-truth global yield stress).
While this result suggests that overfitting was again avoided, the error is about $10\%$ higher than in the surrogate for the mean stresses only.
In the posterior distributions, the uncertainty in $m$, $g_0$, and $\sigma_\text{noise}$ is generally comparable to the calibrations using both mean and grain-average stresses (\autoref{fig:marginals_one_stage}bd).
However, the distributions for $m$, $H$, $g_0 / g_1$, and $\sigma_\text{noise}$ are all significantly biased in \autoref{fig:marginals_one_stage}c, even though the error on the test set did not indicate any serious problems with the surrogate in an average sense.
The Surrogate-All case thereby highlights both the advantages of including local data (e.g., reduced uncertainty) and the challenges of using a surrogate (e.g., possible bias in the model).

\subsubsection*{Posterior Predictive Distributions}
\label{sec:ppd}
The previous sections focused on comparing the posterior distributions for the calibration parameters across the different calibration procedures.
Now, the posterior predictive distributions are compared using simulations with a new microstructure for three sets of posterior distributions: both stages of the two-stage calibration and the Surrogate-All case.
These cases were chosen to study potential limitations arising from using surrogate models and excluding local data from the calibration.

\begin{figure}
	\centering
	\includegraphics[width=0.4\textwidth]{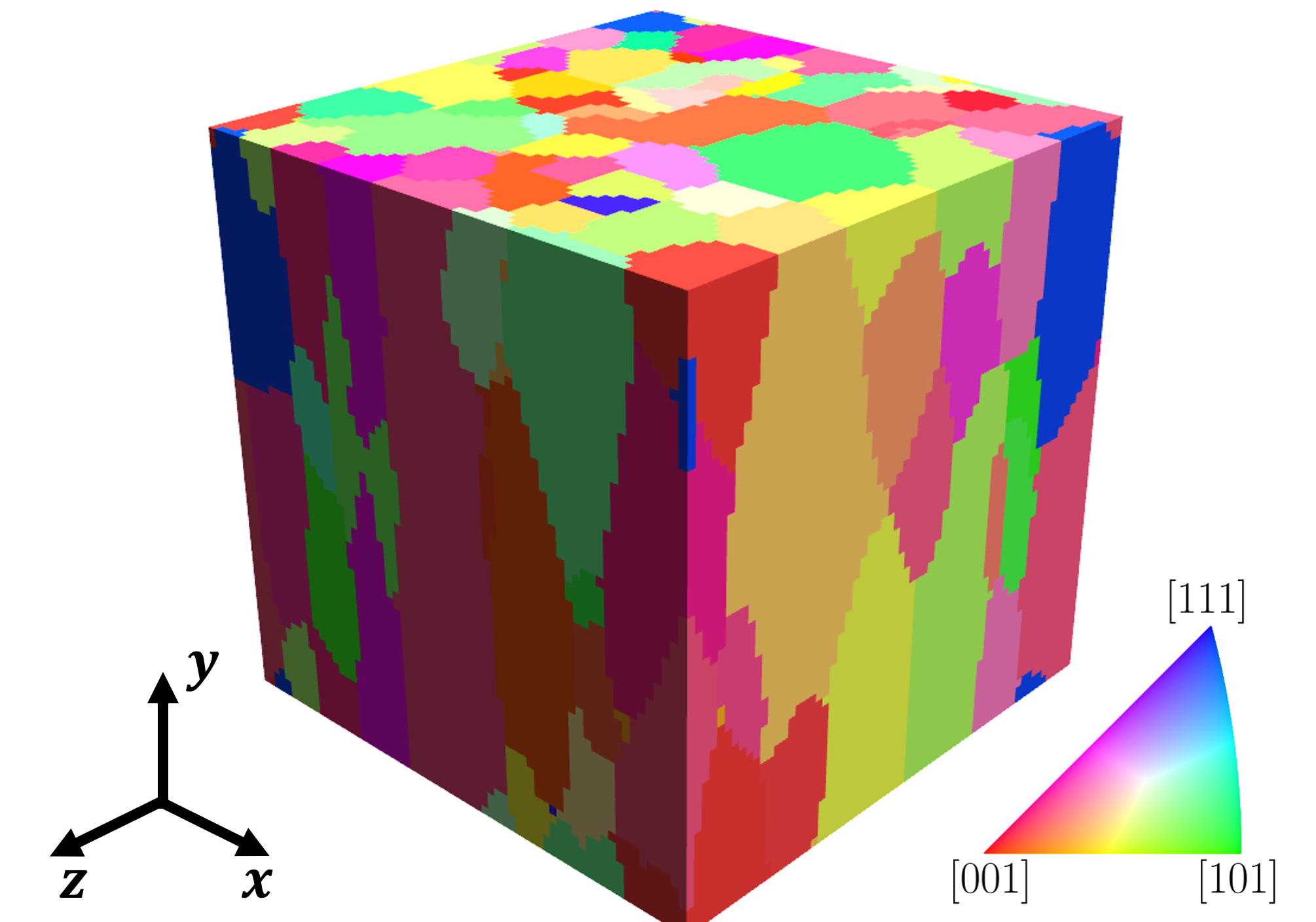}
	\caption{IPF map of the microstructure used for posterior predictive simulations.
	The $y$ direction is the loading direction.}
	\label{fig:new_microstructure}
\end{figure}

The microstructure, shown in \autoref{fig:new_microstructure}, was taken from the periodic statistically-equivalent microstructures in \cite{pribe_multi-model_2025} generated based on EBSD data from additively manufactured Inconel 718.
A constant mean strain rate of $\dot{\gamma}_0 / 10 = 0.0004$ s$^{-1}$ was applied in the $y$ direction up to $1.5\%$ strain.
The strain rate was $90\%$ lower, and the total strain was $50\%$ higher than the corresponding values in the calibration study.
This experiment therefore tests the extrapolation capabilities of the calibrated model.
%

In all cases, the posterior predictive distributions were estimated using forward Monte Carlo simulations to propagate 1288 samples from the posterior distribution through the full-field CP model.
The predictions therefore reveal the influence of bias in the posterior distributions generated using surrogate models (\autoref{fig:two_stage_marginals}a for stage one and \autoref{fig:marginals_one_stage}c for the Surrogate-All case).
The $95\%$ credible interval contains the central $95\%$ of all predicted stresses, representing the uncertainty induced by the propagation of the four calibration parameters.
The $95\%$ predictive interval includes propagation of the estimated noise, such that a future experimental measurement would lie within these bounds with $95\%$ probability.
The ground-truth data was generated by simulating the new microstructure using the ground-truth parameters defined in \autoref{tab:priors}.
%
\begin{figure}
	\centering
	\begin{subfigure}{0.49\textwidth}
		\includegraphics[width=\textwidth]{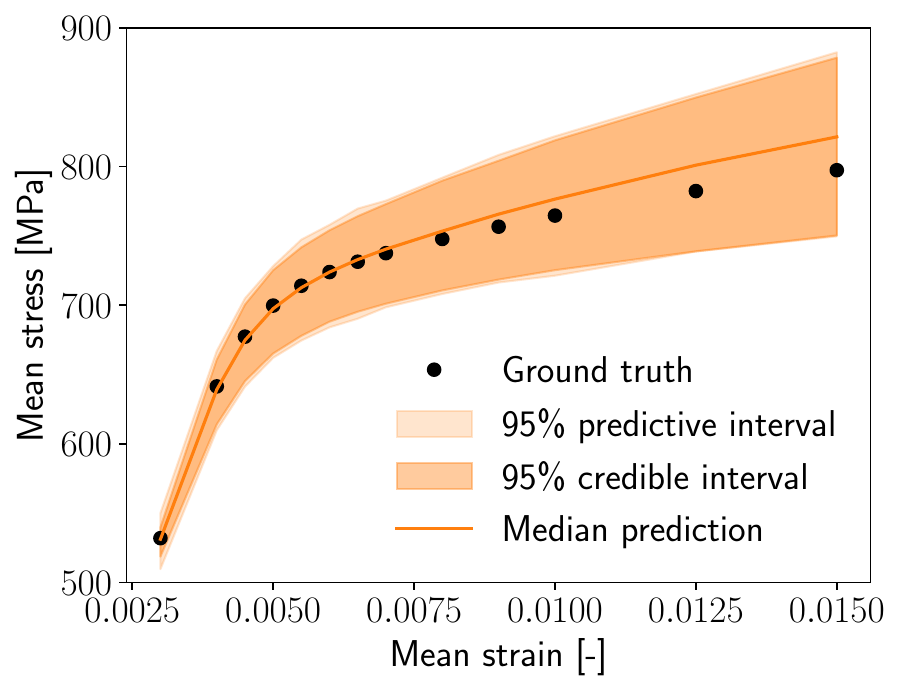}
		\caption{Stage one of the two-stage calibration.}
	\end{subfigure}
	\hfill
	\begin{subfigure}{0.49\textwidth}
		\includegraphics[width=\textwidth]{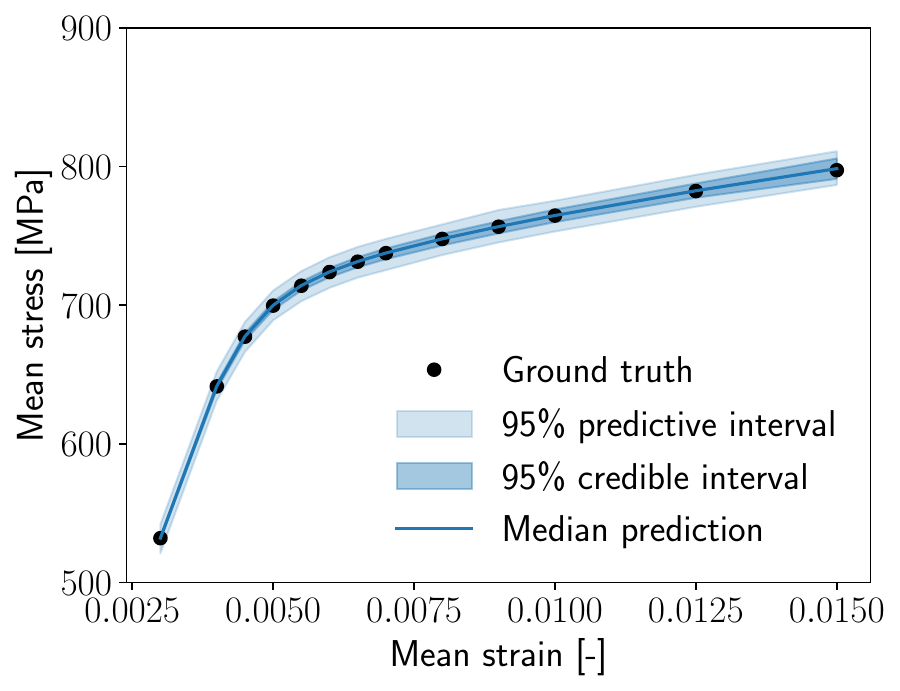}
		\caption{Stage two of the two-stage calibration.}
	\end{subfigure}
    \begin{subfigure}{0.49\textwidth}
		\includegraphics[width=\textwidth]{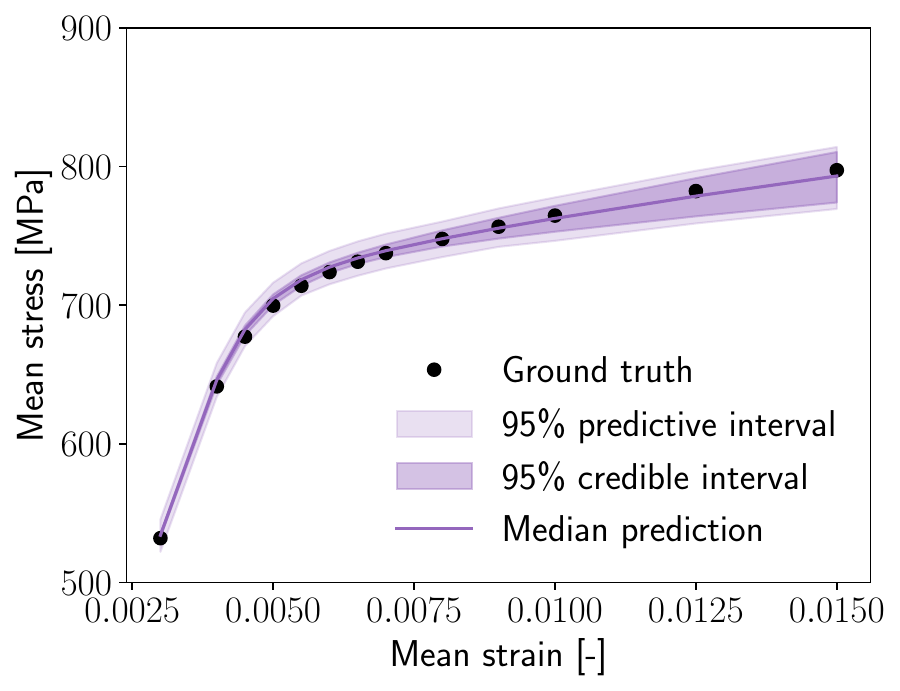}
		\caption{Surrogate model with mean and grain-average stresses (Surrogate-All).}
	\end{subfigure}
	\caption{Posterior predictive distributions for the mean stress with the microstructure in \autoref{fig:new_microstructure}.}
    \label{fig:pp_mean_stress}
\end{figure}

Posterior predictive distributions for the mean stress are shown in \autoref{fig:pp_mean_stress}.
In all three cases, the intervals become wider at mean strains beyond $1\%$, which were not reached in the calibration data.
Stage one has the widest $95\%$ credible intervals, demonstrating the limitations of using only global data, but the intervals include the ground-truth data at all mean strain values.
Stage two closely matches the ground truth and has the smallest $95\%$ credible intervals, reflecting the reduced uncertainty in the posterior distributions compared to stage one.
Finally, the Surrogate-All case also has smaller credible and predictive intervals but reflects the bias observed in \autoref{fig:marginals_one_stage}c for the posterior distributions.
Specifically, the ground-truth stress values for the first four points fall outside the $95\%$ credible interval, suggesting that the model is overconfident in the parameter values.
%
\begin{figure}
	\centering
	\begin{subfigure}{0.49\textwidth}
		\includegraphics[width=\textwidth]{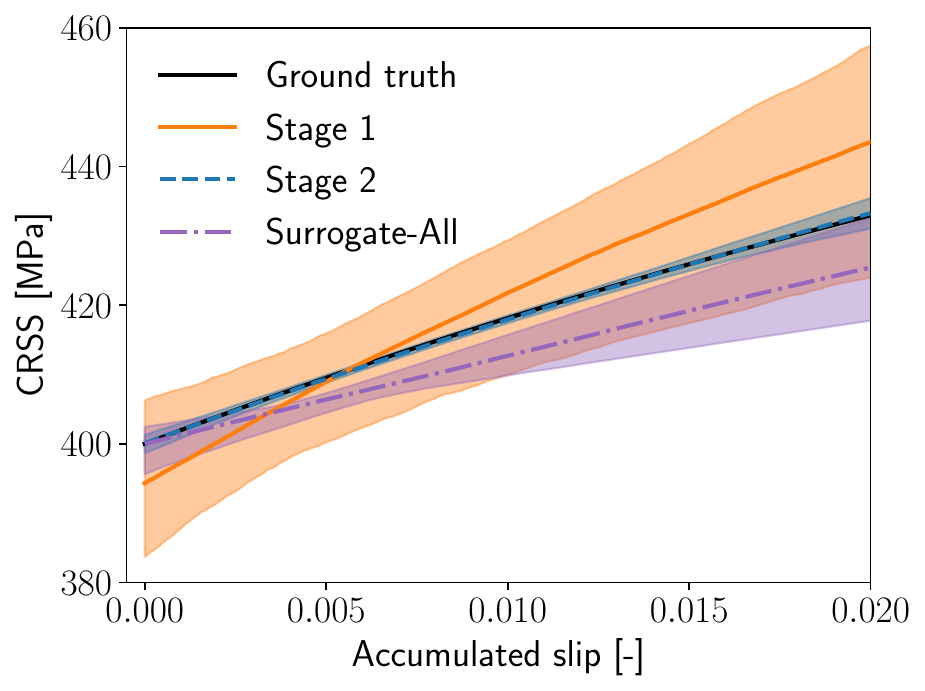}
		\caption{Accumulated slip values up to $0.02$, focusing on initial hardening behavior.}
	\end{subfigure}
	\hfill
	\begin{subfigure}{0.49\textwidth}
		\includegraphics[width=\textwidth]{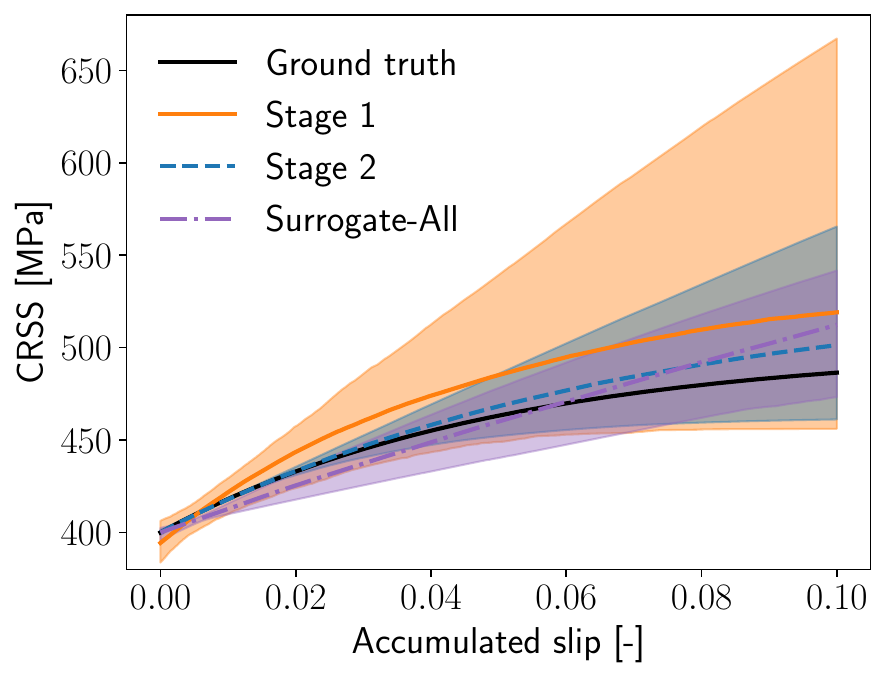}
		\caption{Accumulated slip values up to $0.1$, focusing on extrapolation to higher accumulated slip.}
	\end{subfigure}
	\caption{Posterior predictive distributions for the hardening law. The shaded regions represent the $95\%$ credible intervals, while the lines represent the median predictions.
	Note the different scales on each axis in (a) and (b).}
    \label{fig:pp_hardening}
\end{figure}

$95\%$ credible intervals for the hardening law are shown in \autoref{fig:pp_hardening} to study the ability of each set of calibration parameters to extrapolate to larger deformations.
Specifically, Eq.~\eqref{eq:hardening} was numerically integrated to obtain the CRSS, $g_s$, as a function of the accumulated slip, $\Gamma$, up to $\Gamma = 0.1$ using the same posterior samples as in \autoref{fig:pp_mean_stress}.
From \autoref{fig:pp_hardening}a, the median predictions from all three calibrations are close to the ground truth at low accumulated slip values.
However, the $95\%$ credible interval for the Surrogate-All case does not include the ground-truth values between about $\Gamma = 0.002$ and $\Gamma = 0.02$.
This issue provides further evidence of overconfidence and bias in the surrogate.

From \autoref{fig:pp_hardening}b, the $95\%$ credible intervals become wider at larger accumulated slip values, particularly for stage one, which is consistent with wider $95\%$ credible intervals at larger strains in \autoref{fig:pp_mean_stress}.
For stage two, the median prediction moves away from the ground-truth CRSS values but still follows the shape of the ground-truth hardening curve.
The stage one median prediction also generally follows the shape of the hardening curve, but the $95\%$ credible interval is more than twice as wide as that of stage two.
Meanwhile, bias in the Surrogate-All posterior distributions for $H$ and $g_0 / g_1$ (\autoref{fig:marginals_one_stage}c) causes significant overprediction of the saturation stress, which leads to a nearly linear hardening curve that is inconsistent with the curvature and saturation behavior of the ground truth.

\section*{Discussion}
\label{sec:discussion}
\subsubsection*{Advantages and Limitations of the Two-Stage Calibration}
The two-stage calibration procedure presented in this study was intended to balance the computational efficiency of a surrogate model in stage one with the accuracy of the full CP model in stage two.
In addition, grain-average stresses, mimicking results of an HEDM experiment, were made available in stage two to demonstrate the importance of including local data in the calibration.
All calibrations that included the grain-average stresses have less uncertainty in the posterior distributions, particularly for $m$ and $g_0$, than calibrations using the mean stresses only.
Interestingly, the reduced grain stresses case (\autoref{fig:marginals_comparison}b) showed clear bias in the posterior distributions for $m$ and $g_0$, but this bias was not present in the increased noise case (\autoref{fig:marginals_comparison}c).
Therefore, the quantity of local calibration data may be more important than the precision of the data, although further study is needed on the influence of realistic noise on the local data.
Another interesting result is that including local data significantly reduces uncertainty in the inverse rate sensitivity, $m$, even though the data is only available at a single global strain rate.
Although the global strain rate is constant, each grain sees a different local effective strain rate.

The computational cost for the CP-All case is about twice the cost of stage two in the two-stage calibration.
The number of SMC steps is also higher than for stage two, but only by about 22\%.
The reason for this disparity is that using the posterior from stage one as the stage-two prior regularizes the calibration by reducing the number of computationally-challenging samples in the first few SMC steps, such as samples with high $g_0 / g_1$.
As a result, in the first 72 hours, the CP-All case completed only 8 SMC steps while stage two completed 14 SMC steps.
Thus, the informative priors not only reduced the number of SMC steps but also eliminated parameter combinations in the support of the original priors that require longer CP simulations.
On the other hand, the surrogate model training cost should also be considered for the two-stage calibration.
Assuming 1000 total training simulations and an average of 30 minutes per simulation, 500 CPU hours would be added to the computational cost, which was about 2400 CPU days for the two-stage calibration.
With access to the same 644 cores as the two-step calibration, training would add only one hour of walltime.

The results from the various calibrations in this study also demonstrate potential pitfalls of using surrogate models.
The second mode at low values of $H$ in \autoref{fig:two_stage_marginals}c suggests that the amount of data in stage two may not have been sufficient to counteract the bias in the prior from stage one.
The bias is larger in the posterior distributions for the Surrogate-All case, \autoref{fig:marginals_one_stage}c.
The surrogate is also overconfident in this case: in the posterior predictive distributions, the $95\%$ credible intervals for the Surrogate-All case exclude several of the ground-truth data points.
On the other hand, the $95\%$ credible intervals for the stage-one surrogate are much wider but include all of the ground-truth data.
This wider support in turn allowed stage two to reduce uncertainty and mostly correct the surrogate bias using the full-field CP model and grain-average stress data.
Thus, using only the mean stress calibration data in stage one helped the model to eliminate unrealistic values in the original uniform prior distributions without introducing overconfidence.
More complex constitutive laws and loading conditions are also expected to exacerbate the effects of uncertainty due to using only global data and predictive errors due to biased and overconfident calibrations with a surrogate model.%

\subsubsection*{Implications for Calibration with Real Measurement Data and More Complex Models}
Two important considerations when setting up a CP model are choosing an appropriate constitutive model and identifying the required calibration data.
The results of this study suggest that the portion of the stress-strain curve that is relevant to the intended model application should inform model selection.
It is possible that 1\% mean strain was not sufficiently far into the saturation part of the stress-strain curve to confidently calibrate the hardening parameters $H$ and $g_0 / g_1$.
The posterior predictive distributions for the hardening curve in \autoref{fig:pp_hardening} support this observation and suggest that saturation-type hardening laws are inappropriate if large-strain data is not available or not of interest in the target application.
Nonetheless, the significantly reduced uncertainty in $m$ and $g_0$ when grain-average stresses are included in the calibration demonstrates the value of measuring local data in calibration procedures.

Even though a relatively simple constitutive model was chosen for this study to enable interpretable results, caution was still required in setting up the calibration problem.
To avoid sampling non-physical input combinations, the derived parameter $g_1 / g_0$ was used in the SMC algorithm.
Alternatively, joint priors encoding physical constraints could have been specified for $g_0$, $H$, and $H_d$.
The stage-one surrogate model was useful for identifying this challenge and assessing the physical significance of the calibration parameters.

A challenge for the calibrations in the current study was the correlation between $H$ and $g_0 / g_1$ in \autoref{fig:pairwise_H_ratio}, which highlights the importance of having model parameters that can be uniquely identified.
Interestingly, Sedighiani et al.\ \cite{sedighiani_efficient_2020} obtained a similar result using a genetic algorithm-based calibration with synthetic global uniaxial stress data at several global strain rates.
For a saturation-type hardening law, they obtained a unique parametrization for the initial CRSS and rate sensitivity but not for the hardening parameters, which were also found to be correlated with one another.
The calibration data included strains up to $40\%$, which indicates that the correlation structure and identifiability challenges may persist even with data at higher strains, although no local data was used in the calibration in \cite{sedighiani_efficient_2020}.

In more complex models, additional parameters can provide increased flexibility at the cost of reduced identifiability.
For example, Choi et al.\ \cite{choi_multi-objective_2026} found that a simple Voce hardening law provided better predictions than complicated latent hardening representations whose parameters were difficult to calibrate uniquely.
Correlations across parameters can produce multimodal posterior distributions and significant bias in parameter estimates, especially if the calibration procedure assumes parameter independence \cite{tallman_uncertainty_2020,kucerova_efficient_2023}.
To improve both identifiability and expressiveness, future work could consider combining physics-based model forms like those in \cite{ashraf_robust_2021,dindarlou_substructure-sensitive_2022,ashraf_history_2023,dindarlou_optimization_2024} with the statistical approach in the current study for more complex behavior requiring a phenomenological constitutive description.
Discrepancies due to CP model form error, the microstructural representation, and simplified boundary conditions would also need to be considered in practice.

\begin{figure}
	\centering
	\includegraphics[width=0.5\textwidth]{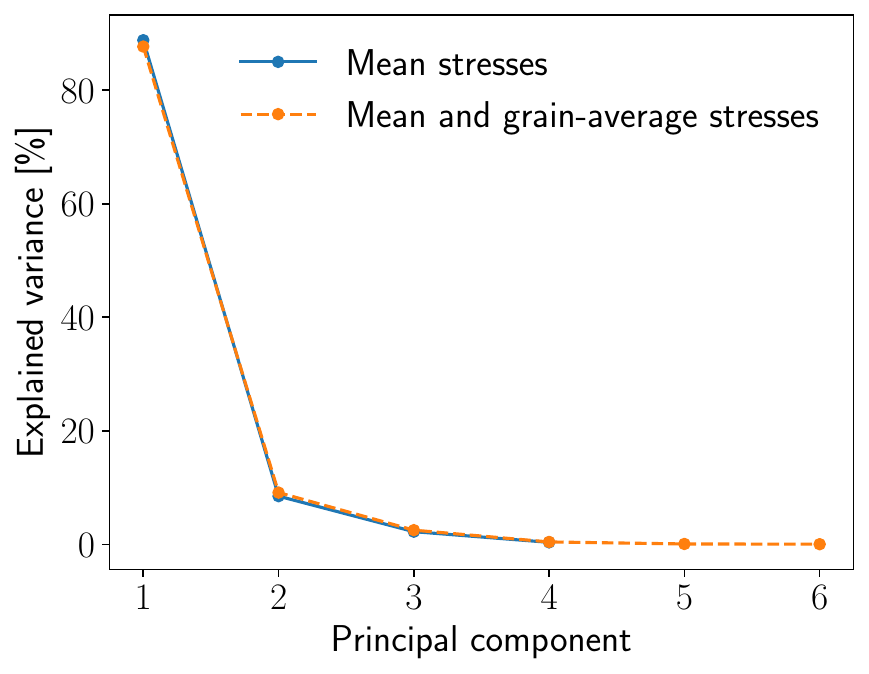}
	\caption{Explained variances, expressed as percentages, for the PCs of the mean stress training data (used for the stage-one surrogate) and the mean and grain-average stress training data (used for Surrogate-All case).}
	\label{fig:explained_variance}
\end{figure}

Another interesting finding related to identifiability comes from the training data for the surrogate models for stage one and the Surrogate-All case.
Both surrogates were trained using the PCs that explained at least $99.9\%$ of the variance in the respective training data sets, resulting in four PCs for the mean stresses and six PCs for the mean and grain-average stresses together.
The explained variances for each PC are shown in \autoref{fig:explained_variance} and are remarkably similar despite the much larger dimensionality of the second dataset.
Thus, the main benefit of the local data here may be simply increasing the amount of calibration data at different effective strain rates, rather than adding knowledge of the physics at smaller length scales.
The grain-average stresses may be more heterogeneous in cases with stronger anisotropy, such as hexagonal close-packed metals.
%

\subsubsection*{Algorithmic Considerations for SMC}
The embarrassingly parallelizable nature of SMC was necessary for completing any calibrations that used the full-field CP model in this study.
As noted in the \nameref{sec:results} section, stage two of the two-stage calibration and the CP-All case required about 2400 days and 4800 days of CPU time, respectively, which is almost certainly intractable for algorithms like MCMC that iteratively select a single chain of points through the parameter space.
Nonetheless, little SMC hyperparameter tuning was performed, so it is likely that improved sampling could be achieved.
For example, the use of five MCMC proposals per SMC step achieved mutation ratios between 0.6 and 0.9, where the mutation ratio represents the proportion of particles that have moved to a new location in the parameter space during the SMC step.
The optimal number of MCMC proposals is the smallest number that achieve a mutation ratio close to 1.0.
The surrogate model could be used to estimate the optimal number of MCMC proposals using multiple SMC runs, assuming that the optimum is similar for the surrogate and the full-field CP model.

The SMC algorithm also likely struggled to sample from the joint distributions of $H$ and $g_0 / g_1$ (\autoref{fig:pairwise_H_ratio}) due to the correlation structure.
These joint distributions are similar to so-called ``banana-shaped distributions'', which are used as a test case for adaptive proposals in MCMC-based algorithms \cite{haario_adaptive_1999}.
The posterior distributions of parameters like $H$ and $g_0 / g_1$ could be sampled more efficiently by guiding the MCMC proposals using estimates of the parameter correlations \cite{sejdinovic_kernel_2014,schuster_kernel_2017}.
Incorporating autodifferentiation into the CP solver \cite{pundir_simplifying_2025} could further improve the MCMC proposals by providing derivatives to the algorithm.

\section*{Summary and Conclusions}
\label{sec:conclusions}
In this study, SMC was used to probabilistically calibrate the parameters of a CP model using Bayesian inference with synthetic global and local measurement data.
A two-stage calibration approach was developed to balance computational cost with accuracy by using a surrogate model in stage one and the full-field CP model in stage two.
Local measurement data in the form of grain-average stresses was added in stage two.
Iterations on the two-stage approach explored robustness of the method to degradation in the local data and compared it with one-stage calibrations.
Analyzing the posterior distributions for the calibration parameters demonstrated the benefits of including local data and revealed causes of uncertainty and bias in the calibration.
Overall, the results support the following conclusions:
\begin{itemize}
	\item SMC enables tractable Bayesian inference with a full-field CP model by using independent, parallel model evaluations to sample from the parameter posterior distributions. Stage two of the two-stage calibration required more than $100,000$ CP simulations and about 2400 days of CPU time but only 3.5 days of wallclock time.
	\item Using informative priors from a calibration with a surrogate model (stage one in the two-stage calibration) further reduces the computational expense by reducing the number of required SMC steps and eliminating non-physical parameter combinations from the support of the prior. On the other hand, bias from the surrogate model in stage one can propagate through to the parameter posterior distributions in stage two, and a one-stage calibration using only the CP model without needing the surrogate at all is also enabled through SMC parallelization.
	\item Including local data (grain-average stresses) in the calibration significantly reduces uncertainty for the initial CRSS and rate sensitivity, producing a corresponding decrease in uncertainty in the posterior predictive distributions. Uncertainty is also somewhat reduced for the hardening parameters, but uniquely identifying these parameters is challenging with the available data. This result highlights the importance of matching the constitutive model, the target application, and the available calibration data.
	\item The reduced uncertainty in the initial CRSS and rate sensitivity is robust to degradation of the local data (i.e., the ``reduced grain stresses'' and ``increased noise'' cases), but reducing the availability of the local data introduces bias to both parameters. This result suggests that the quantity of local data is more important than precision, although further detailed study is needed on this point.
	\item Correlations between parameters can make unique parameter identification difficult or impossible and can arise inherently in the model form or due to limited data. Parameter identifiability should therefore be a consideration in selecting a constitutive law and designing the calibration experiments.
	\item Using a surrogate model with both global and local data in the calibration produced posterior distributions with similar uncertainty but significantly more bias than the two-stage procedure. Surrogate models are thereby useful to identify the type and amount of data that should be used for calibration. However, the resulting calibration may be biased and may lead to poor predictive capability.
\end{itemize}

\subsubsection*{Conflict of Interest Statement}
On behalf of all authors, the corresponding author states that there is no conflict of interest.

\subsubsection*{Data Availability Statement}
The Materialite code used for the crystal plasticity simulations is available at \url{https://github.com/nasa/materialite}. The SMCPy code used in the SMC calibration runs is available at \url{https://github.com/nasa/SMCPy}.

\subsubsection*{Acknowledgements}
The authors gratefully acknowledge support from the Transformational Tools and Technologies Project in the NASA Aeronautics Research Mission Directorate.
The authors also would like to thank Julia Truong, an intern at NASA Langley Research Center in 2025, for her contributions to the SMCPy code.

\begin{appendices}
\section{Fourier-Galerkin solution method}
\label{app:solver}
First, the strain, $\strain$, is expressed as
\begin{equation}
    \label{eq:strain}
    \strain(\mbf{x}) = \mbf{E} + \strain^*(\mbf{x}) \quad \forall \mbf{x} \in \Omega,
\end{equation}
where $\mbf{x}$ is a point in an $\Omega$-periodic domain, $\mbf{E}$ is the mean strain in the domain, and $\strain^*$ is the fluctuation strain, representing the difference between the strain at a point and the mean strain.
The strain field must be compatible, and the stress, $\bm{\sigma}$, which is a nonlinear function of the strain, must satisfy equilibrium:
\begin{equation}
    \nabla \cdot \stress (\mbf{x}) = \mbf{0}, \quad \bm{\sigma(\mbf{x})} = \stress\left(\strain(\mbf{x})\right).
\end{equation}

The weak form for this problem is
\begin{equation}
    \label{eq:weak_form}
    \int_\Omega \delta \strain^*(\mbf{x}) : \stress(\mbf{x}) \mathrm{d}\mbf{x} = 0,
\end{equation}
where $\delta \strain^*$ is a test strain.
Due to periodicity of the strain field, there are no boundary terms in the weak form.
Compatibility of the test strains is enforced using a projection operator, $\mathbb{G}$:
\begin{equation}
    \label{eq:projection}
    \delta \strain^*(\mbf{x}) = \left(\mathbb{G} * \bm{\zeta}\right)(\mbf{x}),
\end{equation}
where $*$ represents convolution and $\bm{\zeta}$ is a second-order tensor field whose compatible part is $\delta \strain^*$.
The operator $\mathbb{G}$ maps second-order tensors to their compatible and symmetric part and has a closed-form expression in frequency space.
This approach does not rely on a reference medium as in other FFT-based methods (e.g., \cite{moulinec_numerical_1998}).
See Appendix \ref{app:projection} for details on the construction of $\mathbb{G}$.
$\mathbb{G}$ is also self-adjoint, allowing the weak form to be written as
\begin{equation}
    \label{eq:weak_form2}
    \int_\Omega \delta \bm{\zeta}(\mbf{x}) : \left(\mathbb{G} * \stress\right)(\mbf{x}) \mathrm{d}\mbf{x} = 0.
\end{equation}

The domain is discretized on a regular grid of voxels, and the stress and strain fields are interpolated using trigonometric polynomial shape functions.
This approach mirrors standard finite element methods, but the nodes and integration points are co-located at the centers of the voxels, and the strain (instead of the displacement) is the primary unknown.
Integrating Eq.~\eqref{eq:weak_form2} using trapezoidal quadrature with the interpolated fields produces the nodal equilibrium equation \cite{zeman_finite_2017}:
\begin{equation}
    \label{eq:equilibrium}
    \ift{\hat{\mathbb{G}}(\bm{\xi}) \ft{\bm{\sigma(\mbf{x})}}} = \mbf{0},
\end{equation}
where $\ft{\cdot}$ represents the Fourier transform, $\hat{\mathbb{G}}$ is the Fourier transform of $\mathbb{G}$, and $\bm{\xi}$ are the frequencies in Fourier space.
To simplify notation on the left-hand side of Eq.~\eqref{eq:equilibrium}, the mapping $\mathcal{G}$ is defined as
\begin{equation}
    \label{eq:G_application}
    \mathcal{G}[\cdot] \coloneq \ift{\hat{\mathbb{G}}(\bm{\xi}) \ft{\cdot}}.
\end{equation}
For nonlinear problems, Eq.~\eqref{eq:equilibrium} is solved iteratively using the Newton-Raphson method.
In iteration $i+1$, the stress field is approximated as
\begin{equation}
    \label{eq:linearization}
    \stress_{(i+1)} \approx \stress_{(i)} + \mathbb{C}_{(i)} \Delta \strain^*_{(i+1)}
\end{equation}
where $\mathbb{C}_{(i)}$ is the consistent tangent operator from iteration $i$, and the dependence on $\mbf{x}$ has been dropped to simplify notation.
Inserting the stress approximation into Eq.~\eqref{eq:equilibrium} gives
\begin{equation}
    \label{eq:nonlinear_equil}
    \mathcal{G} \left[ \mathbb{C}_{(i)} \Delta \strain^*_{(i+1)} \right] = -\mathcal{G}\left[\stress_{(i)}\right]
\end{equation}
Using the FFT, the action of $\mathcal{G}$ can be expressed using a sequence of matrix operations.
As a result, Eq.~\eqref{eq:nonlinear_equil} can be expressed as a linear system of equations, $\mathcal{A} \Delta \strain^*_{(i+1)} = \mbf{b}$, where $\mathcal{A}$ is the system matrix and $\mbf{b}$ is the right-hand side of Eq.~\eqref{eq:nonlinear_equil}.
The system is solved for $\Delta \strain^*_{(i+1)}$ in each iteration using conjugate gradients, which ensures that a compatible strain field is returned in each iteration \cite{mishra_comparative_2016}.
The updated stress and consistent tangent are then obtained from the constitutive model.

Unlike FFT-based methods that rely on a reference medium, the consistent tangent is identical to the one in a typical finite element material subroutine.
Here, the consistent tangent is approximated by differentiating Eqs. \eqref{eq:additive_decomposition} and \eqref{eq:flow}:
\begin{equation}
	\label{eq:tangent}
    \mathbb{C} = \frac{\partial \stress}{\partial \strain} = \left(\mathbb{I}_\text{sym} + \mathbb{D} \frac{\partial\dot{\strain}_p}{\partial \stress} \right)^{-1} \mathbb{D},
\end{equation}
where $\mathbb{I}_\text{sym}$ is the fourth-order symmetric identity tensor, and the unknown derivative is given by
\begin{equation}
	\label{eq:depsp_dsig}
	\frac{\partial \dot{\strain}_p}{\partial \stress} \approx m \dot{\gamma}_0 \sum_{s=1}^{N_s} \frac{\mbf{m}_s \otimes \mbf{m}_s}{g_s} \left( \frac{\mbf{m}_s : \stress}{g_s} \right)^{m-1}.
\end{equation}
Note that the contribution from the hardening law has been neglected in Eq.~\eqref{eq:depsp_dsig}.

\section{Projection tensor and boundary conditions}
\label{app:projection}
For the nonzero, non-Nyquist frequencies, the projection tensor is given in index notation by \cite{zeman_finite_2017}
\begin{equation}
	\hat{\mathbb{G}}(\bm{\xi}) = \hat{G}_{ijkl}(\bm{\xi}) = \frac{1}{2} \frac{ \xi_i \delta_{jk} \xi_l + \xi_i \delta_{jl} \xi_k + \xi_j \delta_{ik} \xi_l + \xi_j \delta_{il} \xi_k}{||\mbf{\xi}||^2} - \frac{\xi_i \xi_j \xi_k \xi_l}{||\mbf{\xi}||^4}.
\end{equation}
Note that $\bm{\xi}$ represents scaled frequencies, with $\xi_i = k_i / L_i$, where $k_i$ is the raw frequencies and $L_i$ is the number of points in the domain in the $i^\text{th}$ direction.
Following \cite{de_geus_finite_2017}, the projection tensor is set to zero at the Nyquist frequencies since the grid used in this study has an even number of points.
Following \cite{lucarini_algorithm_2019}, the components of the projection tensor at the zero frequency are set based on the boundary conditions:
\begin{itemize}
	\item If component $ij$ is under strain control, then $\hat{G}_{ijkl}(\mbf{0})$ is set to zero: $\hat{G}_{ijkl}(\mbf{0}) = 0_{ijkl}$.
	\item If component $ij$ is under stress control, then $\hat{G}_{ijkl}(\mbf{0})$ is set to the corresponding component of the symmetric identity tensor: $\hat{G}_{ijkl}(\mbf{0}) = \frac{1}{2} \left(\delta_{ik} \delta_{jl} + \delta_{il} \delta_{jk} \right)$.
\end{itemize}
For the boundary conditions in this study (uniaxial strain in the $x$ direction, with all other components stress free), the $11$ component is under strain control, and the remaining five components are under stress control.
Thus, the full piecewise definition of the projection tensor is
\begin{equation}
	\hat{G}_{ijkl} (\bm{\xi}) =
	\begin{cases}
		0_{ijkl}, & \text{Nyquist frequencies} \\
		0_{ijkl}, & \bm{\xi} = \mbf{0} \text{ and }ij=11 \text{ (strain control)} \\
		\frac{1}{2} \left(\delta_{ik} \delta_{jl} + \delta_{il} \delta_{jk} \right), & \bm{\xi} = \mbf{0} \text{ and }ij\neq 11 \text{ (stress control)} \\
		\frac{1}{2} \frac{ \xi_i \delta_{jk} \xi_l + \xi_i \delta_{jl} \xi_k + \xi_j \delta_{ik} \xi_l + \xi_j \delta_{il} \xi_k}{||\mbf{\xi}||^2} - \frac{\xi_i \xi_j \xi_k \xi_l}{||\mbf{\xi}||^4}, & \text{otherwise}
	\end{cases}
\end{equation}

\section{PCA for the training data}
\label{app:pca}
To train the neural network surrogate model in the \nameref{sec:sensitivity} section, PCA was conducted on the mean stresses from the 900 training samples.
Specifically, this involves determining the PCs of the matrix containing the 13 mean stress values from each sample:
\begin{equation}
	\label{eq:training_matrix}
	\begin{array}{c@{\hspace{2pt}}c}
		& \text{Mean stresses} \\
		\multirow[c]{1}{*}[18pt]{\rotatebox{-90}{\text{Samples}}} & 
		\overbrace{\begin{bmatrix}
		\Sigma_{11}^{(1,1)} & \Sigma_{11}^{(1,2)} & \cdots & \Sigma_{11}^{(1,13)} \\
		\Sigma_{11}^{(2,1)} & \Sigma_{11}^{(2,2)} & \cdots & \Sigma_{11}^{(2,13)} \\
		\vdots & \vdots & \ddots & \vdots \\
		\Sigma_{11}^{(900,1)} & \Sigma_{11}^{(900,2)} & \cdots & \Sigma_{11}^{(900,13)}
		\end{bmatrix}} \\
	\end{array}
\end{equation}
where each row represents a sample and each column represents a feature such that $\Sigma_{11}^{(i,j)}$ represents the $j^\text{th}$ mean stress for the $i^\text{th}$ sample.
The PCs of this matrix represent the directions of maximum variance in the 13-dimensional feature space.
Two properties of the PCs are useful here:
\begin{enumerate}
	\item Each PC explains a percentage of the variance in the dataset. The first PC explains the most variance, while the $13^\text{th}$ PC explains the least. Choosing the first $M < 13$ PCs reduces the dimensionality of the feature space, where $M$ can be chosen such that a desired percentage of the variance is explained (here, $99.9\%$ of the variance, resulting in $M = 4$).
	\item The PCs are independent, while the original features are highly correlated as inferred from \autoref{fig:pairwise_stress_params}. The independence of the PCs simplified training of the neural network and resulted in more physically realistic stress-strain curve predictions.
\end{enumerate}

For the surrogate model in the \nameref{sec:one_stage} section trained on the mean \emph{and} grain-average stresses from the 900 training samples, the data matrix becomes
\begin{equation}
	\label{eq:training_matrix2}
	\begin{bmatrix}
		\Sigma_{11}^{(1,1)} & \cdots & \Sigma_{11}^{(1,13)} & \sigma_{11,1}^{(1,1)} & \sigma_{11,1}^{(1,2)} & \cdots & \sigma_{11,1}^{(1,13)} & \sigma_{11,2}^{(1,1)} & \cdots & \sigma_{11,32}^{(1,13)} \\
		\Sigma_{11}^{(2,1)} & \cdots & \Sigma_{11}^{(2,13)} & \sigma_{11,1}^{(2,1)} & \sigma_{11,1}^{(2,2)} & \cdots & \sigma_{11,1}^{(2,13)} & \sigma_{11,2}^{(2,1)} & \cdots & \sigma_{11,32}^{(2,13)} \\
		\vdots & \ddots & \vdots & \vdots & \vdots & \ddots & \vdots & \vdots & \ddots & \vdots \\
		\Sigma_{11}^{(900,1)} & \cdots & \Sigma_{11}^{(900,13)} & \sigma_{11,1}^{(900,1)} & \sigma_{11,1}^{(900,2)} & \cdots & \sigma_{11,1}^{(900,13)} & \sigma_{11,2}^{(900,1)} & \cdots & \sigma_{11,32}^{(900,13)}
	\end{bmatrix}
\end{equation}
where $\sigma_{11,m}^{(i,j)}$ represents the $j^\text{th}$ grain-average stress from the $m^\text{th}$ grain for the $i^\text{th}$ sample.
Now, each sample has 429 total features across the mean and grain-average stresses.
Due to the increased dimensionality of the training data, the dimensionality reduction from PCA becomes more powerful, in addition to the benefits of having independent features.

\section{Summary statistics for the marginal posterior distributions}
\label{app:statistics}
\setcounter{table}{0}
The tables in this appendix show six summary statistics for the marginal posterior distributions: maximum a posteriori (MAP) estimate, median, mean, standard deviation, $95\%$ credible interval (CI) centered on the median, and the length of the $95\%$ CI.
The MAP estimates come from a KDE fit to the joint posterior in each case.
The KDE fit may be less reliable for some of the more complex distributions, particularly $H$ and $g_0 / g_1$.
Larger differences between the MAP, median, and mean for these two parameters indicate reduced normality relative to the posterior distributions for $m$, $g_0$, and $\sigma_\text{noise}$, which is also seen visually in the histograms in the \nameref{sec:results} section.
\begin{table}
	\centering
	\caption{Statistics for the posterior distribution of $m$.}
		\begin{tabular}{ccccccc}
		\toprule
		Calibration & MAP & Median & Mean & St.\ dev. & $95\%$ CI & \specialcell{$95\%$ CI\\ length} \\
		\midrule
		\specialcell{Stage one of two-stage\\ calibration (\autoref{fig:two_stage_marginals}a)} & 10.02 & 10.13 & 10.25 & 1.35 & $[7.96, 13.23]$ & 5.27 \\
		\midrule
		\specialcell{Stage two two-stage\\ calibration (\autoref{fig:two_stage_marginals}b)} & 10.05 & 10.02 & 10.00 & 0.126 & $[9.77, 10.25]$ & 0.482 \\
		\midrule
		\specialcell{Reduced grain stresses\\ (\autoref{fig:marginals_comparison}b)} & 9.72 & 9.74 & 9.74 & 0.150 & $[9.46, 10.02]$ & 0.563 \\
		\midrule
		Increased noise (\autoref{fig:marginals_comparison}c) & 9.87 & 9.96 & 9.96 & 0.194 & $[9.60, 10.31]$ & 0.708 \\
		\midrule
		\specialcell{One-stage CP model with\\ mean stresses only (\autoref{fig:marginals_one_stage}a)} & 9.52 & 9.93 & 9.98 & 0.861 & $[8.49, 11.81]$ & 3.32 \\
		\midrule
		\specialcell{One-stage CP model with mean\\ and grain-average stresses (\autoref{fig:marginals_one_stage}b)} & 10.01 & 10.01 & 10.01 & 0.113 & $[9.80, 10.23]$ & 0.435 \\
		\midrule
		\specialcell{One-stage surrogate with mean\\ and grain-average stresses (\autoref{fig:marginals_one_stage}c)} & 10.49 & 10.44 & 10.44 & 0.127 & $[10.20, 10.70]$ & 0.815 \\
		\bottomrule
	\end{tabular}
	\label{tab:posteriors_m}
\end{table}
%
\begin{table}
	\centering
	\caption{Statistics for the posterior distribution of $g_0$. All numbers have units of MPa.}
		\begin{tabular}{ccccccc}
		\toprule
		Calibration & MAP & Median & Mean & St.\ dev. & $95\%$ CI & \specialcell{$95\%$ CI\\ length} \\
		\midrule
		\specialcell{Stage one of two-stage\\ calibration (\autoref{fig:two_stage_marginals}a)} & 394.3 & 394.2 & 394.5 & 5.76 & $[384.0, 407.0]$ & 23.0 \\
		\midrule
		\specialcell{End of two-stage\\ calibration (\autoref{fig:two_stage_marginals}b)} & 399.7 & 400.1 & 400.1 & 0.67 & $[398.7, 401.3]$ & 2.63 \\
		\midrule
		\specialcell{Reduced grain stresses\\ (\autoref{fig:marginals_comparison}b)} & 401.3 & 401.4 & 401.4 & 0.73 & $[400.0, 402.8]$ & 2.81 \\
		\midrule
		Increased noise (\autoref{fig:marginals_comparison}c) & 400.9 & 400.4 & 400.4 & 0.94 & $[398.7, 402.2]$ & 3.49 \\
		\midrule
		\specialcell{One-stage CP model with\\ mean stresses only (\autoref{fig:marginals_one_stage}a)} & 398.2 & 395.8 & 395.9 & 4.08 & $[388.6, 404.2]$ & 15.7 \\
		\midrule
		\specialcell{One-stage CP model with mean\\ and grain-average stresses (\autoref{fig:marginals_one_stage}b)} & 400.1 & 400.1 & 400.1 & 0.64 & $[398.8, 401.3]$ & 2.56 \\
		\midrule
		\specialcell{One-stage surrogate with mean\\ and grain-average stresses (\autoref{fig:marginals_one_stage}c)} & 400.2 & 400.0 & 399.8 & 1.72 & $[395.7, 402.4]$ & 6.74 \\
		\bottomrule
	\end{tabular}
	\label{tab:posteriors_g0}
\end{table}
%
\begin{table}
	\centering
	\caption{Statistics for the posterior distribution of $g_0$. All numbers have units of MPa.}
		\begin{tabular}{ccccccc}
		\toprule
		Calibration & MAP & Median & Mean & St.\ dev. & $95\%$ CI & \specialcell{$95\%$ CI\\ length} \\
		\midrule
		\specialcell{Stage one of two-stage\\ calibration (\autoref{fig:two_stage_marginals}a)} & 9463 & 9937 & 12065 & 8376 & $[1520, 28572]$ & 27052 \\
		\midrule
		\specialcell{End of two-stage\\ calibration (\autoref{fig:two_stage_marginals}b)} & 12311 & 7743 & 8155 & 434.6 & $[2155, 16524]$ & 14369 \\
		\midrule
		\specialcell{Reduced grain stresses\\ (\autoref{fig:marginals_comparison}b)} & 6264 & 5864 & 6825 & 3472 & $[2873, 16239]$ & 13366 \\
		\midrule
		Increased noise (\autoref{fig:marginals_comparison}c) & 4328 & 4116 & 5328 & 2966 & $[2430, 12881]$ & 10452 \\
		\midrule
		\specialcell{One-stage CP model with\\ mean stresses only (\autoref{fig:marginals_one_stage}a)} & 9306 & 10698 & 12189 & 7105 & $[2825, 27713]$ & 24889 \\
		\midrule
		\specialcell{One-stage CP model with mean\\ and grain-average stresses (\autoref{fig:marginals_one_stage}b)} & 8285 & 8306 & 9020 & 3792 & $[3662, 17347]$ & 13685 \\
		\midrule
		\specialcell{One-stage surrogate with mean\\ and grain-average stresses (\autoref{fig:marginals_one_stage}c)} & 1779 & 2101 & 2782 & 2426 & $[1044, 9255]$ & 8211 \\
		\bottomrule
	\end{tabular}
	\label{tab:posteriors_H}
\end{table}
%
\begin{table}
	\centering
	\caption{Statistics for the posterior distribution of $g_0 / g_1$.}
		\begin{tabular}{ccccccc}
		\toprule
		Calibration & MAP & Median & Mean & St.\ dev. & $95\%$ CI & \specialcell{$95\%$ CI\\ length} \\
		\midrule
		\specialcell{Stage one of two-stage\\ calibration (\autoref{fig:two_stage_marginals}a)} & 0.687 & 0.718 & 0.606 & 0.246 & $[0.065, 0.873]$ & 0.808 \\
		\midrule
		\specialcell{End of two-stage\\ calibration (\autoref{fig:two_stage_marginals}b)} & 0.831 & 0.752 & 0.671 & 0.205 & $[0.193, 0.861]$ & 0.668 \\
		\midrule
		\specialcell{Reduced grain stresses\\ (\autoref{fig:marginals_comparison}b)} & 0.713 & 0.707 & 0.692 & 0.121 & $[0.435, 0.879]$ & 0.443 \\
		\midrule
		Increased noise (\autoref{fig:marginals_comparison}c) & 0.585 & 0.572 & 0.574 & 0.170 & $[0.272, 0.838]$ & 0.566 \\
		\midrule
		\specialcell{One-stage CP model with\\ mean stresses only (\autoref{fig:marginals_one_stage}a)} & 0.718 & 0.737 & 0.685 & 0.177 & $[0.200, 0.887]$ & 0.687 \\
		\midrule
		\specialcell{One-stage CP model with mean\\ and grain-average stresses (\autoref{fig:marginals_one_stage}b)} & 0.763 & 0.769 & 0.749 & 0.092 & $[0.531, 0.870]$ & 0.339 \\
		\midrule
		\specialcell{One-stage surrogate with mean\\ and grain-average stresses (\autoref{fig:marginals_one_stage}c)} & 0.333 & 0.378 & 0.415 & 0.145 & $[0.241, 0.806]$ & 0.565 \\
		\bottomrule
	\end{tabular}
	\label{tab:posteriors_ratio}
\end{table}
%
\begin{table}
	\centering
	\caption{Statistics for the posterior distribution of $\sigma_\text{noise}$. All numbers have units of MPa.}
		\begin{tabular}{ccccccc}
		\toprule
		Calibration & MAP & Median & Mean & St.\ dev. & $95\%$ CI & \specialcell{$95\%$ CI\\ length} \\
		\midrule
		\specialcell{Stage one of two-stage\\ calibration (\autoref{fig:two_stage_marginals}a)} & 6.24 & 7.32 & 7.72 & 2.12 & $[4.82, 13.0]$ & 8.18 \\
		\midrule
		\specialcell{End of two-stage\\ calibration (\autoref{fig:two_stage_marginals}b)} & 5.22 & 5.24 & 5.23 & 0.151 & $[4.91, 5.52]$ & 0.618 \\
		\midrule
		\specialcell{Reduced grain stresses\\ (\autoref{fig:marginals_comparison}b)} & 5.16 & 5.18 & 5.20 & 0.251 & $[4.74, 5.72]$ & 0.984 \\
		\midrule
		Increased noise (\autoref{fig:marginals_comparison}c) & 6.92 & 6.85 & 6.97 & 1.29 & $[4.90, 9.78]$ & 4.88 \\
		\midrule
		\specialcell{One-stage CP model with\\ mean stresses only (\autoref{fig:marginals_one_stage}a)} & 5.69 & 6.61 & 6.88 & 1.72 & $[4.42, 11.0]$ & 6.55 \\
		\midrule
		\specialcell{One-stage CP model with mean\\ and grain-average stresses (\autoref{fig:marginals_one_stage}b)} & 5.34 & 5.24 & 5.24 & 0.163 & $[4.92, 5.55]$ & 0.622 \\
		\midrule
		\specialcell{One-stage surrogate with mean\\ and grain-average stresses (\autoref{fig:marginals_one_stage}c)} & 5.98 & 5.98 & 5.98 & 0.208 & $[5.59, 6.41]$ & 0.815 \\
		\bottomrule
	\end{tabular}
	\label{tab:posteriors_sigma}
\end{table}
\end{appendices}

\clearpage
\singlespacing
\bibliography{refs}

@article{tallman_uncertainty_2020,
	title = {Uncertainty propagation in reduced order models based on crystal plasticity},
	volume = {365},
	doi = {10.1016/j.cma.2020.113009},
	language = {en},
	urldate = {2021-10-01},
	journal = {Comput Methods Appl Mech Eng},
	author = {Tallman, Aaron E. and Swiler, Laura P. and Wang, Yan and McDowell, David L.},
	month = jun,
	year = {2020},
	pages = {113009},
}

@article{moulinec_numerical_1998,
	title = {A numerical method for computing the overall response of nonlinear composites with complex microstructure},
	volume = {157},
	doi = {10.1016/S0045-7825(97)00218-1},
	language = {en},
	number = {1-2},
	urldate = {2021-10-21},
	journal = {Comput Methods Appl Mech Eng},
	author = {Moulinec, H. and Suquet, P.},
	year = {1998},
	pages = {69--94},
}

@incollection{ghosh_non-deterministic_2020,
	address = {Cham},
	title = {Non-deterministic calibration of crystal plasticity model parameters},
	language = {en},
	urldate = {2021-11-02},
	booktitle = {Integrated {Computational} {Materials} {Engineering} ({ICME})},
	publisher = {Springer International Publishing},
	author = {Hochhalter, Jacob and Bomarito, Geoffrey and Yeratapally, Saikumar and Leser, Patrick and Ruggles, Tim and Warner, James and Leser, William},
	editor = {Ghosh, Somnath and Woodward, Christopher and Przybyla, Craig},
	year = {2020},
	pages = {165--198},
}

@article{tari_validation_2018,
	title = {Validation of micro-mechanical {FFT}-based simulations using {High} {Energy} {Diffraction} {Microscopy} on {Ti}-{7Al}},
	volume = {154},
	doi = {10.1016/j.actamat.2018.05.036},
	language = {en},
	urldate = {2021-11-17},
	journal = {Acta Mater},
	author = {Tari, Vahid and Lebensohn, Ricardo A. and Pokharel, Reeju and Turner, Todd J. and Shade, Paul A. and Bernier, Joel V. and Rollett, Anthony D.},
	year = {2018},
	pages = {273--283},
}

@article{mcdowell_microstructure-sensitive_2010,
	title = {Microstructure-sensitive computational modeling of fatigue crack formation},
	volume = {32},
	doi = {10.1016/j.ijfatigue.2010.01.003},
	language = {en},
	number = {9},
	urldate = {2022-03-18},
	journal = {Int J Fatigue},
	author = {McDowell, D.L. and Dunne, F.P.E.},
	year = {2010},
	pages = {1521--1542},
}

@article{leser_sequential_2018,
	title = {Sequential {Monte} {Carlo}: {Enabling} real-time and high-fidelity prognostics},
	volume = {10},
	shorttitle = {Sequential {Monte} {Carlo}},
	doi = {10.36001/phmconf.2018.v10i1.564},
	number = {1},
	urldate = {2022-04-14},
	journal = {Annual Conference of the PHM Society},
	author = {Leser, Patrick E. and Hochhalter, Jacob D. and Warner, James E and Bomarito, Geoffrey F. and Leser, William P. and Yuan, Fuh-Gwo},
	year = {2018},
}

@article{brynjarsdottir_learning_2014,
	title = {Learning about physical parameters: the importance of model discrepancy},
	volume = {30},
	doi = {10.1088/0266-5611/30/11/114007},
	number = {11},
	urldate = {2022-04-22},
	journal = {Inverse Problems},
	author = {Brynjarsd{\'o}ttir, Jenn{\'y} and O'Hagan, Anthony},
	year = {2014},
	pages = {114007},
}

@article{whelan_uncertainty_2019,
	title = {Uncertainty quantification in {ICME} workflows for fatigue critical computational modeling},
	volume = {220},
	doi = {10.1016/j.engfracmech.2019.106673},
	language = {en},
	urldate = {2022-06-28},
	journal = {Eng Fract Mech},
	author = {Whelan, Gary and McDowell, David L.},
	year = {2019},
	pages = {106673},
}

@article{rovinelli_assessing_2017,
	title = {Assessing reliability of fatigue indicator parameters for small crack growth via a probabilistic framework},
	volume = {25},
	doi = {10.1088/1361-651X/aa6c45},
	number = {4},
	urldate = {2022-06-28},
	journal = {Modelling Simul Mater Sci Eng},
	author = {Rovinelli, Andrea and Guilhem, Yoann and Proudhon, Henry and Lebensohn, Ricardo A and Ludwig, Wolfgang and Sangid, Michael D},
	year = {2017},
	pages = {045010},
}

@article{pagan_modeling_2017,
	title = {Modeling slip system strength evolution in {Ti}-{7Al} informed by in-situ grain stress measurements},
	volume = {128},
	doi = {10.1016/j.actamat.2017.02.042},
	journal = {Acta Mater},
	author = {Pagan, Darren C. and Shade, Paul A. and Barton, Nathan R. and Park, Jun-Sang and Kenesei, Peter and Menasche, David B. and Bernier, Joel V.},
	year = {2017},
	pages = {406--417},
}

@article{zhang_crack_2016,
	title = {Crack nucleation using combined crystal plasticity modelling, high-resolution digital image correlation and high-resolution electron backscatter diffraction in a superalloy containing non-metallic inclusions under fatigue},
	volume = {472},
	doi = {10.1098/rspa.2015.0792},
	language = {en},
	number = {2189},
	urldate = {2022-09-08},
	journal = {Proc R Soc A},
	author = {Zhang, Tiantian and Jiang, Jun and Britton, Ben and Shollock, Barbara and Dunne, Fionn},
	year = {2016},
	pages = {20150792},
}

@article{venkatraman_bayesian_2022,
	title = {Bayesian analysis of parametric uncertainties and model form probabilities for two different crystal plasticity models of lamellar grains in $\alpha$+$\beta$ {Titanium} alloys},
	volume = {154},
	doi = {10.1016/j.ijplas.2022.103289},
	language = {en},
	urldate = {2022-09-08},
	journal = {Int J Plast},
	author = {Venkatraman, Aditya and McDowell, David L. and Kalidindi, Surya R.},
	year = {2022},
	pages = {103289},
}

@article{prithivirajan_direct_2021,
	title = {Direct comparison of microstructure-sensitive fatigue crack initiation via crystal plasticity simulations and in situ high-energy {X}-ray experiments},
	volume = {197},
	doi = {10.1016/j.matdes.2020.109216},
	language = {en},
	urldate = {2022-09-15},
	journal = {Mater Des},
	author = {Prithivirajan, Veerappan and Ravi, Priya and Naragani, Diwakar and Sangid, Michael D.},
	year = {2021},
	pages = {109216},
}

@article{bandyopadhyay_uncertainty_2019,
	title = {Uncertainty quantification in the mechanical response of crystal plasticity simulations},
	volume = {71},
	doi = {10.1007/s11837-019-03551-3},
	language = {en},
	number = {8},
	urldate = {2022-11-10},
	journal = {JOM},
	author = {Bandyopadhyay, Ritwik and Prithivirajan, Veerappan and Sangid, Michael D.},
	year = {2019},
	pages = {2612--2624},
}

@article{del_moral_sequential_2006,
	title = {Sequential monte carlo samplers},
	volume = {68},
	doi = {10.1111/j.1467-9868.2006.00553.x},
	number = {3},
	journal = {J R Stat Soc Ser B Stat Methodol},
	author = {Del Moral, Pierre and Doucet, Arnaud and Jasra, Ajay},
	year = {2006},
	pages = {411--436},
}

@article{del_moral_adaptive_2012,
	title = {On adaptive resampling strategies for sequential {Monte} {Carlo} methods},
	volume = {18},
	doi = {10.3150/10-BEJ335},
	number = {1},
	urldate = {2022-11-29},
	journal = {Bernoulli},
	author = {Del Moral, Pierre and Doucet, Arnaud and Jasra, Ajay},
	year = {2012},
}

@article{lebensohn_spectral_2020,
	title = {Spectral methods for full-field micromechanical modelling of polycrystalline materials},
	volume = {173},
	doi = {10.1016/j.commatsci.2019.109336},
	language = {en},
	urldate = {2022-12-12},
	journal = {Comput Mater Sci},
	author = {Lebensohn, Ricardo A. and Rollett, Anthony D.},
	year = {2020},
	pages = {109336},
}

@article{menasche_afrl_2021,
	title = {{AFRL} {Additive} {Manufacturing} {Modeling} {Series}: {Challenge} 4, in situ mechanical test of an {IN625} sample with concurrent high-energy diffraction microscopy characterization},
	volume = {10},
	doi = {10.1007/s40192-021-00218-3},
	language = {en},
	number = {3},
	urldate = {2022-12-21},
	journal = {Integr Mater Manuf Innov},
	author = {Menasche, David B. and Musinski, William D. and Obstalecki, Mark and Shah, Megna N. and Donegan, Sean P. and Bernier, Joel V. and Kenesei, Peter and Park, Jun-Sang and Shade, Paul A.},
	year = {2021},
	pages = {338--347},
}

@inproceedings{sangid_validation_2014,
	address = {National Harbor, Maryland},
	title = {Validation of microstructure-based materials modeling},
	doi = {10.2514/6.2014-0462},
	language = {en},
	urldate = {2023-02-07},
	booktitle = {55th {AIAA}/{ASME}/{ASCE}/{AHS}/{ASC} {Structures}, {Structural} {Dynamics}, and {Materials} {Conference}},
	publisher = {American Institute of Aeronautics and Astronautics},
	author = {Sangid, Michael and Yeratapally, Saikumar Reddy and Rovinelli, Andrea},
	year = {2014},
}

@article{stinville_multi-modal_2022,
	title = {Multi-modal dataset of a polycrystalline metallic material: {3D} microstructure and deformation fields},
	volume = {9},
	doi = {10.1038/s41597-022-01525-w},
	language = {en},
	number = {1},
	urldate = {2023-04-06},
	journal = {Sci Data},
	author = {Stinville, J. C. and Hestroffer, J. M. and Charpagne, M. A. and Polonsky, A. T. and Echlin, M. P. and Torbet, C. J. and Valle, V. and Nygren, K. E. and Miller, M. P. and Klaas, O. and Loghin, A. and Beyerlein, I. J. and Pollock, T. M.},
	year = {2022},
	pages = {460},
}

@article{dindarlou_substructure-sensitive_2022,
	title = {Substructure-sensitive crystal plasticity with material-invariant parameters},
	volume = {155},
	doi = {10.1016/j.ijplas.2022.103306},
	language = {en},
	urldate = {2023-06-19},
	journal = {Int J Plast},
	author = {Dindarlou, Shahram and Castelluccio, Gustavo M.},
	year = {2022},
	pages = {103306},
}

@article{mackey_grain_2023,
	title = {Grain interactions under thermo-mechanical loads investigated with coupled crystal plasticity simulations and high-energy {X}-ray diffraction microscopy},
	volume = {257},
	doi = {10.1016/j.actamat.2023.119166},
	language = {en},
	urldate = {2023-09-05},
	journal = {Acta Mater},
	author = {Mackey, Brandon T. and Bandyopadhyay, Ritwik and Gustafson, Sven E. and Sangid, Michael D.},
	year = {2023},
	pages = {119166},
}

@article{menasche_four-dimensional_2023,
	title = {Four-dimensional microstructurally small fatigue crack growth in cyclically loaded nickel superalloy specimen},
	volume = {177},
	doi = {10.1016/j.ijfatigue.2023.107920},
	journal = {Int J Fatigue},
	author = {Menasche, David B. and Shade, Paul A. and Kenesei, Peter and Park, Jun-Sang and Musinski, William D.},
	year = {2023},
	pages = {107920},
}

@article{yeratapally_discrepancy_2021,
	title = {Discrepancy between crystal plasticity simulations and far-field high-energy {X}-ray diffraction microscopy measurements},
	volume = {10},
	doi = {10.1007/s40192-021-00216-5},
	number = {2},
	journal = {Integr Mater Manuf Innov},
	author = {Yeratapally, Saikumar R. and Cerrone, Albert R. and Glaessgen, Edward H.},
	year = {2021},
	pages = {196--217},
}

@article{pedregosa_scikit-learn_2011,
	title = {Scikit-learn: {Machine} learning in {Python}},
	volume = {12},
	journal = {Journal of Machine Learning Research},
	author = {Pedregosa, F. and Varoquaux, G. and Gramfort, A. and Michel, V. and Thirion, B. and Grisel, O. and Blondel, M. and Prettenhofer, P. and Weiss, R. and Dubourg, V. and Vanderplas, J. and Passos, A. and Cournapeau, D. and Brucher, M. and Perrot, M. and Duchesnay, E.},
	year = {2011},
	pages = {2825--2830},
}

@article{ghorbanpour_crystal_2017,
	title = {A crystal plasticity model incorporating the effects of precipitates in superalloys: {Application} to tensile, compressive, and cyclic deformation of {Inconel} 718},
	volume = {99},
	doi = {10.1016/j.ijplas.2017.09.006},
	journal = {Int J Plast},
	author = {Ghorbanpour, Saeede and Zecevic, Milovan and Kumar, Anil and Jahedi, Mohammad and Bicknell, Jonathan and Jorgensen, Luke and Beyerlein, Irene J. and Knezevic, Marko},
	year = {2017},
	pages = {162--185},
}

@article{zeman_finite_2017,
	title = {A finite element perspective on nonlinear {FFT}-based micromechanical simulations},
	volume = {111},
	doi = {10.1002/nme.5481},
	number = {10},
	journal = {Int J Numer Meth Engr},
	author = {Zeman, J. and de Geus, T. W. J. and Vond{\v r}ejc, J. and Peerlings, R. H. J. and Geers, M. G. D.},
	year = {2017},
	pages = {903--926},
}

@article{de_geus_finite_2017,
	title = {Finite strain {FFT}-based non-linear solvers made simple},
	volume = {318},
	doi = {10.1016/j.cma.2016.12.032},
	journal = {Comput Methods Appl Mech Eng},
	author = {de Geus, T. W. J. and Vond{\v r}ejc, J. and Zeman, J. and Peerlings, R. H. J. and Geers, M. G. D.},
	year = {2017},
	pages = {412--430},
}

@article{lucarini_algorithm_2019,
	title = {An algorithm for stress and mixed control in {Galerkin}-based {FFT} homogenization},
	volume = {119},
	doi = {10.1002/nme.6069},
	number = {8},
	journal = {Int J Numer Meth Engr},
	author = {Lucarini, S. and Segurado, J.},
	year = {2019},
	pages = {797--805},
}

@article{nguyen_bayesian_2021,
	title = {Bayesian calibration of a physics-based crystal plasticity and damage model},
	volume = {149},
	doi = {10.1016/j.jmps.2020.104284},
	journal = {J Mech Phys Solids},
	author = {Nguyen, Thao and Francom, Devin C. and Luscher, D. J. and Wilkerson, J. W.},
	year = {2021},
	pages = {104284},
}

@article{tran_asynchronous_2023,
	title = {An asynchronous parallel high-throughput model calibration framework for crystal plasticity finite element constitutive models},
	volume = {72},
	doi = {10.1007/s00466-023-02308-9},
	number = {3},
	journal = {Comput Mech},
	author = {Tran, Anh and Lim, Hojun},
	year = {2023},
	pages = {485--498},
}

@article{depriester_crystal_2023,
	title = {Crystal plasticity simulations of in situ tensile tests: {A} two-step inverse method for identification of {CP} parameters, and assessment of {CPFEM} capabilities},
	volume = {168},
	doi = {10.1016/j.ijplas.2023.103695},
	journal = {Int J Plast},
	author = {Depriester, D. and Goulmy, J. P. and Barrallier, L.},
	year = {2023},
	pages = {103695},
}

@article{veasna_machine_2023,
	title = {Machine learning-based multi-objective optimization for efficient identification of crystal plasticity model parameters},
	volume = {403},
	doi = {10.1016/j.cma.2022.115740},
	journal = {Comput Methods Appl Mech Eng},
	author = {Veasna, Khem and Feng, Zhangxi and Zhang, Qi and Knezevic, Marko},
	year = {2023},
	pages = {115740},
}

@article{sangid_coupling_2020,
	title = {Coupling in situ experiments and modeling {\textendash} {Opportunities} for data fusion, machine learning, and discovery of emergent behavior},
	volume = {24},
	doi = {10.1016/j.cossms.2019.100797},
	number = {1},
	journal = {Curr Opin Solid State Mater Sci},
	author = {Sangid, Michael D.},
	year = {2020},
	pages = {100797},
}

@article{dindarlou_optimization_2024,
	title = {Optimization of crystal plasticity parameters with proxy materials data for alloy single crystals},
	volume = {174},
	doi = {10.1016/j.ijplas.2024.103894},
	journal = {Int J Plast},
	author = {Dindarlou, Shahram and Castelluccio, Gustavo M.},
	year = {2024},
	pages = {103894},
}

@article{pribe_multi-model_2025,
	title = {Multi-model {Monte} {Carlo} estimation for crystal plasticity structure{\textendash}property simulations of additively manufactured metals},
	volume = {247},
	doi = {10.1016/j.commatsci.2024.113481},
	journal = {Comput Mater Sci},
	author = {Pribe, Joshua D. and Leser, Patrick E. and Yeratapally, Saikumar R. and Glaessgen, Edward H.},
	year = {2025},
	pages = {113481},
}

@article{pundir_simplifying_2025,
	title = {Simplifying {FFT}-based methods for solid mechanics with automatic differentiation},
	volume = {435},
	doi = {10.1016/j.cma.2024.117572},
	language = {en},
	urldate = {2025-01-14},
	journal = {Comput Methods Appl Mech Eng},
	author = {Pundir, Mohit and Kammer, David S.},
	year = {2025},
	pages = {117572},
}

@article{mishra_comparative_2016,
	title = {A comparative study on low-memory iterative solvers for {FFT}-based homogenization of periodic media},
	volume = {321},
	doi = {10.1016/j.jcp.2016.05.041},
	language = {en},
	urldate = {2025-01-14},
	journal = {J Comput Phys},
	author = {Mishra, Nachiketa and Vond{\v r}ejc, Jaroslav and Zeman, Jan},
	year = {2016},
	pages = {151--168},
}

@article{venkatraman_bayesian_2025,
	title = {Bayesian protocols for high-throughput identification of kinematic hardening model forms},
	volume = {285},
	doi = {10.1016/j.ijmecsci.2024.109791},
	language = {en},
	urldate = {2025-02-11},
	journal = {Int J Mech Sci},
	author = {Venkatraman, Aditya and Johnson, Camilla E. and McDowell, David L. and Kalidindi, Surya R.},
	year = {2025},
	pages = {109791},
}

@article{lim_quantitative_2025,
	title = {Quantitative comparison between experiments and crystal plasticity simulations using microstructural clones},
	volume = {184},
	doi = {10.1016/j.ijplas.2024.104186},
	language = {en},
	urldate = {2025-08-18},
	journal = {Int J Plast},
	author = {Lim, Hojun and Fitzgerald, Kaitlynn M. and Ruggles, Timothy J. and Gilliland, William G. and Aragon, Nicole K. and Carroll, Jay D.},
	year = {2025},
	pages = {104186},
}

@article{shimanek_simultaneous_2025,
	title = {Simultaneous optimization of crystal plasticity hardening parameters},
	volume = {77},
	doi = {10.1007/s11837-024-06935-2},
	language = {en},
	number = {1},
	urldate = {2025-08-18},
	journal = {JOM},
	author = {Shimanek, John D. and Liu, Zi-Kui and Beese, Allison M.},
	year = {2025},
	pages = {324--335},
}

@article{shade_exploring_2019,
	title = {Exploring new links between crystal plasticity models and high-energy {X}-ray diffraction microscopy},
	volume = {23},
	doi = {10.1016/j.cossms.2019.07.002},
	language = {en},
	number = {5},
	urldate = {2025-08-18},
	journal = {Curr Opin Solid State Mater Sci},
	author = {Shade, Paul A. and Musinski, William D. and Obstalecki, Mark and Pagan, Darren C. and Beaudoin, Armand J. and Bernier, Joel V. and Turner, Todd J.},
	year = {2019},
	pages = {100763},
}

@article{savage_identification_2021,
	title = {Identification of crystal plasticity model parameters by multi-objective optimization integrating microstructural evolution and mechanical data},
	volume = {379},
	doi = {10.1016/j.cma.2021.113747},
	language = {en},
	urldate = {2025-08-20},
	journal = {Comput Methods Appl Mech Eng},
	author = {Savage, Daniel J. and Feng, Zhangxi and Knezevic, Marko},
	year = {2021},
	pages = {113747},
}

@article{herrera-solaz_inverse_2014,
	title = {An inverse optimization strategy to determine single crystal mechanical behavior from polycrystal tests: {Application} to {AZ31} {Mg} alloy},
	volume = {57},
	doi = {10.1016/j.ijplas.2014.02.001},
	language = {en},
	urldate = {2025-08-21},
	journal = {Int J Plast},
	author = {Herrera-Solaz, V. and LLorca, J. and Dogan, E. and Karaman, I. and Segurado, J.},
	year = {2014},
	pages = {1--15},
}

@article{cheng_macro-micro_2024,
	title = {A macro-micro approach for identifying crystal plasticity parameters for necking and failure in nickel-based alloy {Haynes} 282},
	volume = {178},
	doi = {10.1016/j.ijplas.2024.103997},
	language = {en},
	urldate = {2025-08-21},
	journal = {Int J Plast},
	author = {Cheng, Jiahao and Hu, Xiaohua and Lach, Timothy and Chen, Xiang (Frank)},
	year = {2024},
	pages = {103997},
}

@article{liu_strategy_2020,
	title = {A strategy for synthetic microstructure generation and crystal plasticity parameter calibration of fine-grain-structured dual-phase steel},
	volume = {126},
	doi = {10.1016/j.ijplas.2019.10.002},
	language = {en},
	urldate = {2025-08-21},
	journal = {Int J Plast},
	author = {Liu, Wenqi and Lian, Junhe and Aravas, Nikolaos and M{\"u}nstermann, Sebastian},
	year = {2020},
	pages = {102614},
}

@article{park_crystal_2019,
	title = {Crystal plasticity modeling of 3rd generation multi-phase {AHSS} with martensitic transformation},
	volume = {120},
	doi = {10.1016/j.ijplas.2019.03.010},
	language = {en},
	urldate = {2025-08-21},
	journal = {Int J Plast},
	author = {Park, Taejoon and Hector, Louis G. and Hu, Xiaohua and Abu-Farha, Fadi and Fellinger, Michael R. and Kim, Hyunki and Esmaeilpour, Rasoul and Pourboghrat, Farhang},
	year = {2019},
	pages = {1--46},
}

@article{fernandez-zelaia_creep_2022,
	title = {Creep anisotropy modeling and uncertainty quantification of an additively manufactured {Ni}-based superalloy},
	volume = {151},
	doi = {10.1016/j.ijplas.2021.103177},
	language = {en},
	urldate = {2025-08-21},
	journal = {Int J Plast},
	author = {Fernandez-Zelaia, Patxi and Lee, Yousub and Dryepondt, Sebastien and Kirka, Michael M.},
	year = {2022},
	pages = {103177},
}

@article{kucerova_efficient_2023,
	title = {Efficient probabilistic multi-fidelity calibration of a damage-plastic model for confined concrete},
	volume = {412},
	doi = {10.1016/j.cma.2023.116099},
	language = {en},
	urldate = {2025-08-21},
	journal = {Comput Methods Appl Mech Eng},
	author = {Ku{\v c}erov{\'a}, Anna and S{\'y}kora, Jan and Havl{\'a}sek, Petr and Jaru{\v s}kov{\'a}, Daniela and Jir{\'a}sek, Milan},
	year = {2023},
	pages = {116099},
}

@article{kuhn_identifying_2022,
	title = {Identifying material parameters in crystal plasticity by {Bayesian} optimization},
	volume = {23},
	doi = {10.1007/s11081-021-09663-7},
	language = {en},
	number = {3},
	urldate = {2025-09-15},
	journal = {Optim Eng},
	author = {Kuhn, Jannick and Spitz, Jonathan and Sonnweber-Ribic, Petra and Schneider, Matti and B{\"o}hlke, Thomas},
	year = {2022},
	pages = {1489--1523},
}

@article{haario_adaptive_1999,
	title = {Adaptive proposal distribution for random walk {Metropolis} algorithm},
	volume = {14},
	doi = {10.1007/s001800050022},
	language = {en},
	number = {3},
	urldate = {2025-11-10},
	journal = {Comput Stat},
	author = {Haario, Heikki and Saksman, Eero and Tamminen, Johanna},
	year = {1999},
	pages = {375--395},
}

@article{ashraf_history_2023,
	title = {History and temperature dependent cyclic crystal plasticity model with material-invariant parameters},
	volume = {161},
	doi = {10.1016/j.ijplas.2022.103494},
	language = {en},
	urldate = {2025-11-11},
	journal = {Int J Plast},
	author = {Ashraf, Farhan and Castelluccio, Gustavo M.},
	year = {2023},
	pages = {103494},
}

@article{roters_overview_2010,
	title = {Overview of constitutive laws, kinematics, homogenization and multiscale methods in crystal plasticity finite-element modeling: {Theory}, experiments, applications},
	volume = {58},
	copyright = {https://www.elsevier.com/tdm/userlicense/1.0/},
	doi = {10.1016/j.actamat.2009.10.058},
	language = {en},
	number = {4},
	urldate = {2025-11-14},
	journal = {Acta Mater},
	author = {Roters, F. and Eisenlohr, P. and Hantcherli, L. and Tjahjanto, D.D. and Bieler, T.R. and Raabe, D.},
	year = {2010},
	pages = {1152--1211},
}

@article{pokharel_polycrystal_2014,
	title = {Polycrystal plasticity: {Comparison} between grain-scale observations of deformation and simulations},
	volume = {5},
	doi = {10.1146/annurev-conmatphys-031113-133846},
	language = {en},
	number = {1},
	urldate = {2025-11-14},
	journal = {Annu Rev Condens Matter Phys},
	author = {Pokharel, Reeju and Lind, Jonathan and Kanjarla, Anand K. and Lebensohn, Ricardo A. and Li, Shiu Fai and Kenesei, Peter and Suter, Robert M. and Rollett, Anthony D.},
	year = {2014},
	pages = {317--346},
}

@article{obstalecki_quantitative_2014,
	title = {Quantitative analysis of crystal scale deformation heterogeneity during cyclic plasticity using high-energy {X}-ray diffraction and finite-element simulation},
	volume = {75},
	doi = {10.1016/j.actamat.2014.04.059},
	language = {en},
	urldate = {2025-11-14},
	journal = {Acta Mater},
	author = {Obstalecki, Mark and Wong, Su Leen and Dawson, Paul R. and Miller, Matthew P.},
	year = {2014},
	pages = {259--272},
}

@article{sedighiani_efficient_2020,
	title = {An efficient and robust approach to determine material parameters of crystal plasticity constitutive laws from macro-scale stress{\textendash}strain curves},
	volume = {134},
	doi = {10.1016/j.ijplas.2020.102779},
	language = {en},
	urldate = {2025-11-17},
	journal = {Int J Plast},
	author = {Sedighiani, K. and Diehl, M. and Traka, K. and Roters, F. and Sietsma, J. and Raabe, D.},
	year = {2020},
	pages = {102779},
}

@article{ashraf_robust_2021,
	title = {A robust approach to parameterize dislocation glide energy barriers in {FCC} metals and alloys},
	volume = {56},
	doi = {10.1007/s10853-021-06376-1},
	language = {en},
	number = {29},
	urldate = {2025-11-18},
	journal = {J Mater Sci},
	author = {Ashraf, Farhan and Castelluccio, Gustavo M.},
	year = {2021},
	pages = {16491--16509},
}

@inproceedings{schuster_kernel_2017,
	title = {Kernel sequential {Monte} {Carlo}},
	booktitle = {Joint {European} {Conference} on {Machine} {Learning} and {Knowledge} {Discovery} in {Databases}},
	publisher = {Springer},
	author = {Schuster, Ingmar and Strathmann, Heiko and Paige, Brooks and Sejdinovic, Dino},
	year = {2017},
	pages = {390--409},
}

@inproceedings{sejdinovic_kernel_2014,
	title = {Kernel adaptive {Metropolis}-{Hastings}},
	booktitle = {International {Conference} on {Machine} {Learning}},
	publisher = {PMLR},
	author = {Sejdinovic, Dino and Strathmann, Heiko and Garcia, Maria Lomeli and Andrieu, Christophe and Gretton, Arthur},
	year = {2014},
	pages = {1665--1673},
}

@article{weber_machine_2022,
	title = {Machine learning-enabled self-consistent parametrically-upscaled crystal plasticity model for {Ni}-based superalloys},
	volume = {402},
	doi = {10.1016/j.cma.2022.115384},
	language = {en},
	urldate = {2025-12-11},
	journal = {Comput Methods Appl Mech Eng},
	author = {Weber, George and Pinz, Maxwell and Ghosh, Somnath},
	year = {2022},
	pages = {115384},
}

@article{gilliland_concurrent_2026,
	title = {Concurrent in situ high-resolution electron backscatter diffraction and digital image correlation for full-field stress-strain},
	volume = {32},
	issn = {1431-9276, 1435-8115},
	doi = {10.1093/mam/ozaf130},
	language = {en},
	number = {1},
	urldate = {2026-03-12},
	journal = {Microsc Microanal},
	author = {Gilliland, Will and Ruggles, Tim and Fitzgerald, Kaitlynn and Carroll, Jay and Bomarito, Geoffrey and Hochhalter, Jacob},
	month = jan,
	year = {2026},
	pages = {ozaf130},
}

@article{rezwan_coupled_2025,
	title = {Coupled phase field damage and crystal plasticity analysis of intragranular fracture: {The} role of crystallographic orientation and voids},
	volume = {191},
	issn = {07496419},
	doi = {10.1016/j.ijplas.2025.104372},
	language = {en},
	urldate = {2026-03-12},
	journal = {Int J Plast},
	author = {Rezwan, Aashique A. and Aragon, Nicole K. and De Oca Zapiain, David Montes and Lim, Hojun},
	year = {2025},
	pages = {104372},
}

@article{joshi_integrated_2025,
	title = {Integrated experiment and simulation co-design: {A} key infrastructure for predictive mesoscale materials modeling},
	volume = {211},
	issn = {01676636},
	doi = {10.1016/j.mechmat.2025.105480},
	language = {en},
	urldate = {2026-03-12},
	journal = {Mechanics of Materials},
	author = {Joshi, Shailendra and Bucsek, Ashley and Pagan, Darren C. and Daly, Samantha and Ravindran, Suraj and Marian, Jaime and Bessa, Miguel A. and Kalidindi, Surya R. and Admal, Nikhil C. and Reina, Celia and Ghosh, Somnath and Warren, James A. and Vi{\~n}als, Jorge and Tadmor, Ellad B.},
	month = dec,
	year = {2025},
	pages = {105480},
}

@article{sahu_micro-mechanisms_2024,
	title = {Micro-mechanisms of anisotropic deformation in the presence of notch in commercially pure {Titanium}: {An} in-situ study with {CPFEM} simulations},
	volume = {177},
	issn = {07496419},
	doi = {10.1016/j.ijplas.2024.103985},
	language = {en},
	urldate = {2026-03-12},
	journal = {Int J Plast},
	author = {Sahu, Vivek Kumar and Chakraborty, Pritam and Yadava, Manasij and Gurao, Nilesh P.},
	year = {2024},
	pages = {103985},
}

@article{stinville_insights_2023,
	title = {Insights into plastic localization by crystallographic slip from emerging experimental and numerical approaches},
	volume = {53},
	copyright = {http://creativecommons.org/licenses/by/4.0/},
	doi = {10.1146/annurev-matsci-080921-102621},
	language = {en},
	number = {1},
	urldate = {2026-03-13},
	journal = {Annu Rev Mater Res},
	author = {Stinville, J. C. and Charpagne, M. A. and Maa{\ss}, R. and Proudhon, H. and Ludwig, W. and Callahan, P. G. and Wang, F. and Beyerlein, I. J. and Echlin, M. P. and Pollock, T. M.},
	year = {2023},
	pages = {275--317},
}

@article{proudhon_situ_2020,
	title = {In situ {4D} mechanical testing of structural materials: {The} data challenge},
	volume = {24},
	issn = {13590286},
	shorttitle = {In situ {4D} mechanical testing of structural materials},
	doi = {10.1016/j.cossms.2020.100834},
	language = {en},
	number = {4},
	urldate = {2026-03-13},
	journal = {Curr Opin Solid State Mater Sci},
	author = {Proudhon, Henry and Pelerin, Maxime and King, Andrew and Ludwig, Wolfgang},
	year = {2020},
	pages = {100834},
}

@article{cardoso_coelho_efficient_2024,
	title = {Efficient constitutive parameter identification through optimisation-based techniques: {A} {Comparative} analysis and novel composite {Bayesian} optimisation strategy},
	volume = {427},
	issn = {00457825},
	shorttitle = {Efficient constitutive parameter identification through optimisation-based techniques},
	doi = {10.1016/j.cma.2024.117039},
	language = {en},
	urldate = {2026-05-20},
	journal = {Comput Methods Appl Mech Eng},
	author = {Cardoso Coelho, R.P. and Carvalho Alves, A. Francisca and Andrade Pires, F.M.},
	year = {2024},
	pages = {117039},
}

@article{choi_multi-objective_2026,
	title = {Multi-objective surrogate-assisted calibration of {CPFEM} models using macroscopic response and in situ {EBSD} measurements of grain reorientation trajectories},
	volume = {305},
	doi = {10.1016/j.actamat.2025.121809},
	language = {en},
	urldate = {2026-06-05},
	journal = {Acta Mater},
	author = {Choi, J. and Mur{\'a}nsky, O. and Messner, M.C. and Wei, T. and Hu, T. and Kruzic, J.J. and McMurtrey, M.D.},
	year = {2026},
	pages = {121809},
}
\end{document}